\documentclass[12pt]{iopart}

\usepackage{iopams}
\usepackage{amsmath, amssymb, amstext}
\usepackage{physics}
\usepackage{cite}
\usepackage{caption, subcaption}
\usepackage{latexsym}
\usepackage{graphicx}
\usepackage{hyperref}
\usepackage{color}
\usepackage{enumerate}
\usepackage{tabularx}
\usepackage{floatrow}
\usepackage[normalem]{ulem}

\definecolor{bluc}{cmyk}{1,1,0,0.1}
\definecolor{rossoCP3}{cmyk}{0,.88,.77,.40}
\definecolor{rosso}{cmyk}{0,1,1,0.4}
\definecolor{rossos}{cmyk}{0,1,1,0.55}
\definecolor{rossoc}{cmyk}{0,1,1,0.2}
\definecolor{verdes}{cmyk}{0.92,0,0.59,0.4}
\hypersetup{colorlinks, bookmarksopen, bookmarksnumbered, citecolor=verdes, linkcolor=bluc, pdfstartview=FitH, urlcolor=rossos}

\usepackage{etoolbox}
\makeatletter
\newrobustcmd{\fixappendix}{%
  \patchcmd{\l@section}{1.5em}{7em}{}{}%
  \patchcmd{\l@subsection}{2.3em}{7em}{}{}%
}
\makeatother

\begin{document}

\title[Total light bending in non-asymptotically flat black hole spacetimes]{Total light bending in non-asymptotically flat black hole spacetimes}

\author{Flavio C. Sánchez$^1$, Armando A. Roque$^1$, Benito Rodríguez$^1$, Javier Chagoya$^1$}
\address{$^1$ Unidad Acad\'emica de F\'isica, Universidad Aut\'onoma de Zacatecas, 98060, M\'exico.}
\eads{\mailto{flavio.sanchez@fisica.uaz.edu.mx}, \mailto{arestrada@fisica.uaz.edu.mx}, \mailto{javier.chagoya@fisica.uaz.edu.mx}, \mailto{benito.rodriguez@fisica.uaz.edu.mx}}
\vspace{1.cm}
\begin{abstract}
\noindent 

The gravitational deflection of light is a critical test of modified theories of gravity. A few years ago, Gibbons and Werner introduced a definition of the deflection angle based on the Gauss-Bonnet theorem. In more recent years, Arakida proposed a related idea for defining the deflection angle in non-asymptotically flat spacetimes. We revisit this idea and use it to compute the angular difference in the Kottler geometry and a non-asymptotically flat solution in Horndeski gravity.
Our analytic and numerical calculations show that a triangular array of laser beams can be designed so that the proposed definition of the deflection angle is sensitive to different sources of curvature. Moreover, we find that near the photon sphere, the deflection angle in the Horndeski solution is similar to its Schwarzschild counterpart, and we confirm that the shadows seen by a static observer are identical. 
\end{abstract}

\submitto{\CQG}

\maketitle

\tableofcontents
\clearpage

\section{Introduction}
The study of light has been of great relevance in shaping our understanding of nature~\cite{Zubairy2016}, and in several periods of history, it has been closely related to the study of geometry. In fact, the deflection of light due to the Sun played an important role in establishing General Relativity (GR) as the most accepted gravitational model. According to this theory, the deflection angle of light by the Sun should be twice that predicted by Newton's laws of motion. In $1919$, Eddington and his collaborators successfully confirmed this prediction~\cite{Eddington1920, Earman1980}. Since then, angular deflection has been one of the main tools to study various astrophysical scenarios (see Ref.~\cite{Ellis2010GravitationalLA} for a review).


Indeed, the weak deflection of light by the Sun is one of the classic tests that validate GR~\cite{Shapiro:2004zz}, forcing any other theory of gravity to reduce to GR on solar system scales. However, for observations on larger scales, GR is in the spotlight. This is mainly due to the necessity to introduce a dark sector, i.e., dark matter and dark energy. At large scales, one consequence of the well-known weak lensing effect, which has been used to analyze the validity of various alternative theories of gravity~\cite{Schmidt_2008, Uzan:2010ri, Pratten_2016, Baker:2019gxo}, is the angular deflection. A similar effect, but in the strong gravity regime (beyond the weak field limit), is the strong lensing effect, which provides an experimental test for compact astrophysical objects, such as black holes, e.g., the one embedded in M87 galaxy~\cite{EventHorizonTelescope:2019ggy} and SgrA*~\cite{SagittariusA}, or exotic ultracompact objects like boson stars~\cite{Cunha:2017wao, Liebling:2012fv}, $\ell$-boson stars~\cite{Alcubierre:2018ahf, Alcubierre:2021psa, Alcubierre:2022rgp, Roque:2023sjl}, Horndeski stars~\cite{Barranco:2021auj, Roque:2021lvr}, or other exotic compact objects such as superspinars, anisotropic stars, etc. (see for example table 1 in~\cite{Cardoso:2019rvt}).

The pedagogical methodology for calculating the bending angle of light assumes that the spacetime is static and spherically symmetric and that the light source and the observer are in the flat region at spatial infinity. The calculation for this case is relatively straightforward (see, e.g., Refs.~\cite{Misner:1973prb, Weinberg:1972kfs, Perlick:2010zh}). However, when some of these assumptions are removed, the calculations usually require the use of new approximations or analytical and numerical methods. The angular deflection for the Schwarzschild spacetime was derived by Darwin in $1959$~\cite{doi:10.1098/rspa.1959.0015}. A few years later, using geometrical optics techniques, the bending angle in the Kerr spacetime was studied~\cite{1972ApJ...173L.137C}. In $1978$, a pioneering work was presented by Luminet~\cite{1979A&A....75..228L} to study the optical appearance of a Schwarzschild black hole surrounded by an accretion disk. Recent works examine the angular deflection in more complex scenarios such as Kerr black hole, see, e.g., Refs.~\cite{Bozza:2005tg, Aazami:2011tu, Ghosh:2022mka}, or using perturbative methods~\cite{Bozza_2001, Bozza_2002}. This phenomenon is also studied in the context of modifications to GR~\cite{Virbhadra:1998dy, Izmailov:2019uhy, Chagoya:2020bqz}. Nonetheless, in each of these instances, the presumption of both the light source and the observer being located in an asymptotically flat spacetime remains unchanged. 

In scenarios where the observer or the light sources are in a non-asymptotically flat spacetime, it is not entirely clear how to correctly perform the calculation of the deflection angle. In the context of GR, a solution that is non-asymptotically flat is the Kottler spacetime~\cite{Kottler1918, weyl1919statischen}~\footnote{The Kottler solution is also known as Schwarzschild–(anti)de Sitter solution in dependence on $\Lambda$'s sign.}, where an initial work~\cite{Islam:1983rxp} claimed that the angular deflection is not affected at all by a cosmological constant $\Lambda$. However, a recent analysis proposed by Rindler and Ishak (RI)~\cite{Rindler:2007zz} concludes that the presence of $\Lambda$ diminishes the bending angle.\footnote{The Rindler and Ishak idea has also been applied to other non-asymptotically flat solutions.} See, for example, the references~\cite{Bhattacharya:2009rv, Bhattacharya:2010xh}. Although other works such as~\cite{Hu:2021yzn, Bessa:2022sdh} indicate that the effect would be too small to be observed. Nevertheless, using a different approach, Ref.~\cite{Lake:2001fj} concludes that $\Lambda$ does not produce changes in the bending of light.  

These discussions motivated the search for an alternative formalism to calculate and interpret the deflection angle in curved spacetimes. In Ref.~\cite{Gibbons:2008rj} Gibbons and Werner presented a novel method to compute it by applying the Gauss-Bonnet theorem to the optical metric, for the Schwarzschild spacetime the authors concluded that in the weak field limit, the well-known result for the deflection angle is recovered. In a later work~\cite{Werner:2012rc}, Werner extended the results to the Kerr solution arriving at the same conclusion. The Gibbons and Werner methodology has been applied in various scenarios~\cite{Crisnejo_2018, Ovgun:2018fnk, Jusifi_2017.2}. However, the original methodology cannot be applied when the light source and observer are at a finite distance. In~\cite{Ishihara:2016vdc, Ishihara:2016sfv}, the authors extended the methodology to address these scenarios, opening up the possibility to study non-asymptotically flat spacetimes. This extension allows for the consideration of light sources and observers that are not necessarily located at infinite spatial distances (see Refs.~\cite{Takizawa:2020egm, Takizawa:2020dja} for further studies on this topic).

Recently, based on the extension presented in~\cite{Ishihara:2016vdc, Ishihara:2016sfv}, H. Arakida proposed an alternative approach for calculating the deflection angle of light in a static, spherically symmetric, and non-asymptotically flat spacetime, as described in Refs.~\cite{Arakida:2017hrm, Arakida:2020xil}. Arakida's methodology involves computing the total angular deflection by comparing the sums of internal angles of two polygons: one embedded in the spacetime under study and the other in a background spacetime with zero mass. His work primarily focused on investigating the influence of $\Lambda$ on the angular deflection for the Kottler solution. In this paper, we reframe Arakida's concept as an \emph{angular difference} and explore the feasibility of using triangular configurations as an observational tool.

The goal of this paper is to provide a systematic study of the angular difference. In section~\ref{Sec:II} a description of the theoretical framework underlying the determination of the angular difference is provided. The basic ideas of the Gauss-Bonnet theorem are presented in subsection~\ref{Sec:II.1}, while the Arakida proposal is explained and deduced in subsection~\ref{Sec:II.2}. In subsections~\ref{Sec:II.3} and~\ref{Sec:II.4}, iterative solutions for the angular difference are obtained, first within the framework of GR for the Kottler solution, and then for a spacetime of  Horndeski gravity, which is used as an example of a non-asymptotically flat spacetime that differs from Kottler in some aspects, for instance, its scalar curvature is not constant, and only one of the metric components diverges for large $r$. In both cases, the 
iterative solutions reduce to the Schwarzschild formula for light deflection in a specific limit.

Section~\ref{Sec:III} focuses on the numerical analysis of the angular difference in the Kottler and Horndeski spacetimes. Subsection~\ref{Sec:III.1} provides a detailed explanation of our numerical implementation, e.g., the criteria, and the boundary conditions used for constructing the triangular configurations and performing the numerical calculation of the angular difference. The main results and discussions for the Kottler solution are presented in subsection~\ref{Sec:III.2}, and for the Horndeski solution in subsection~\ref{Sec:III.3}. In both cases, we identify the contributions to the angular  difference coming from the mass and from the other sources of curvature, and we compare the numerical and iterative solutions in order to assess the validity of the latter. Section~\ref{Sec:IV} is devoted to studying the angular radius of the shadows of these solutions. For the Kottler case, we present the known results, while for the Horndeski solution, applying the conventional methodology we show that the angular radius is the same as for a Schwarzschild black hole.  Conclusions, observational possibilities, and open questions are provided in section~\ref{Sec:VI}. Technical results, which include terms not given explicitly in some equations of the main text, are included in the appendix.

{\bf Conventions.---} 
We use the $(-,+,+,+)$ signature for the spacetime metric, and the definitions $R_{\mu\nu\rho}{}^{\sigma}:= \partial_\nu\Gamma^{\sigma}_{\mu\rho}+\Gamma^{\alpha}_{\mu\rho}\Gamma^{\sigma}_{\alpha\nu}-(\nu\leftrightarrow \mu)$, $R_{\mu\nu}:= R_{\mu\alpha\nu}{}^{\alpha}$ for the Riemann and Ricci tensors, and $R:= R_{\mu}{}^{\mu}$ for the Ricci scalar, according to Wald's notation~\cite{Wald:1984rg}. Greek indices on a tensor represent spacetime components, i.e., $\mu = 0, 1, 2, 3$, and Latin indices  refer to spatial components $1, 2, 3$. We work in natural units $\hbar=c=1$. We use Newton's constant $G=6.674\times10^{-11}\; \text{m}^3 \text{kg}^{-1} s^{-2}$, the cosmological constant value $\Lambda=1.1056\times 10^{-52} \text{m}^{-2}$, the Sun's mass, $\mathrm{M}_\odot = 1.9885\times10^{30} \;\text{kg}$, and radius $\mathrm{R}_\odot = 6.96\times10^{8} \; \text{m}$.

\section{Theoretical setup}\label{Sec:II}
In the first part of this section, we describe the geometrical setup underlying the novel concept~\cite{Arakida:2017hrm, Arakida:2020xil} for the deflection angle and its derivation, which is discussed in the second part of the section. In the third part, we explore some analytical approaches to this formulation in the context of GR and a Horndeski extension.

\subsection{Geometrical setup}\label{Sec:II.1}
We assume a static and spherically symmetric spacetime with line element
\begin{align}
    ds^{2}=-h(r)dt^{2}+ f(r)^{-1}dr^{2}+r^2 d\theta^{2}+ r^2 \sin^{2}\theta d\phi ^{2}, \label{metric}
\end{align}
 where the functions $h(r)$ and $f(r)$ are real-valued and depend only on the areal radial coordinate $r$.\footnote{The original works~\cite{Arakida:2017hrm, Arakida:2020xil} focused on exploring the propagation of light in Schwarzschild-like spacetimes. For this reason, the author adopted the particular case $f(r)={h(r)}$ of the metric~(\ref{metric}).} The Riemannian geometry experienced by a light ray is determined by the null condition $ds^{2}=0$ and is defined by the \emph{optical metric} with line element

\begin{align}
dt^{2}&=\bar{g}_{ij}dx^{i}dx^{j}.\label{Eq:lenOptMe}
\end{align}
This metric defines a $3-$dimensional Riemannian space $\mathcal{M}^{3d}_{\text{opt}}$, in which the light ray is described as a spatial curve with affine parameter $t$ and whose unit tangent vector can be defined as
\begin{align}
k^{i}:=\frac{dx^{i}}{dt}.
\label{Eq:UnitSpatVect}
\end{align}
The non-vanishing components of $\bar{g}_{ij}$ are	
\begin{equation}\label{eq:optmetric}
    \bar{g}_{rr}=\frac{1}{f(r) h(r)}, 
 \ \ \ \bar{g}_{\theta\theta}=\frac{r^2}{h(r)}, \ \ \ \bar{g}_{\phi\phi}=\frac{r^2 \sin^2\theta}{h(r)}.
\end{equation}	
For a hypersurface of constant time, these are equivalent to the spatial components of the spacetime metric~(\ref{metric}) after a conformal rescaling by a factor $\omega^2(r)=1/h(r)$. As a result, the angles remain the same in both metrics~\cite{Arakida:2020xil}.

\begin{figure*}[t]
	\centering	
    \includegraphics[width=15.cm]{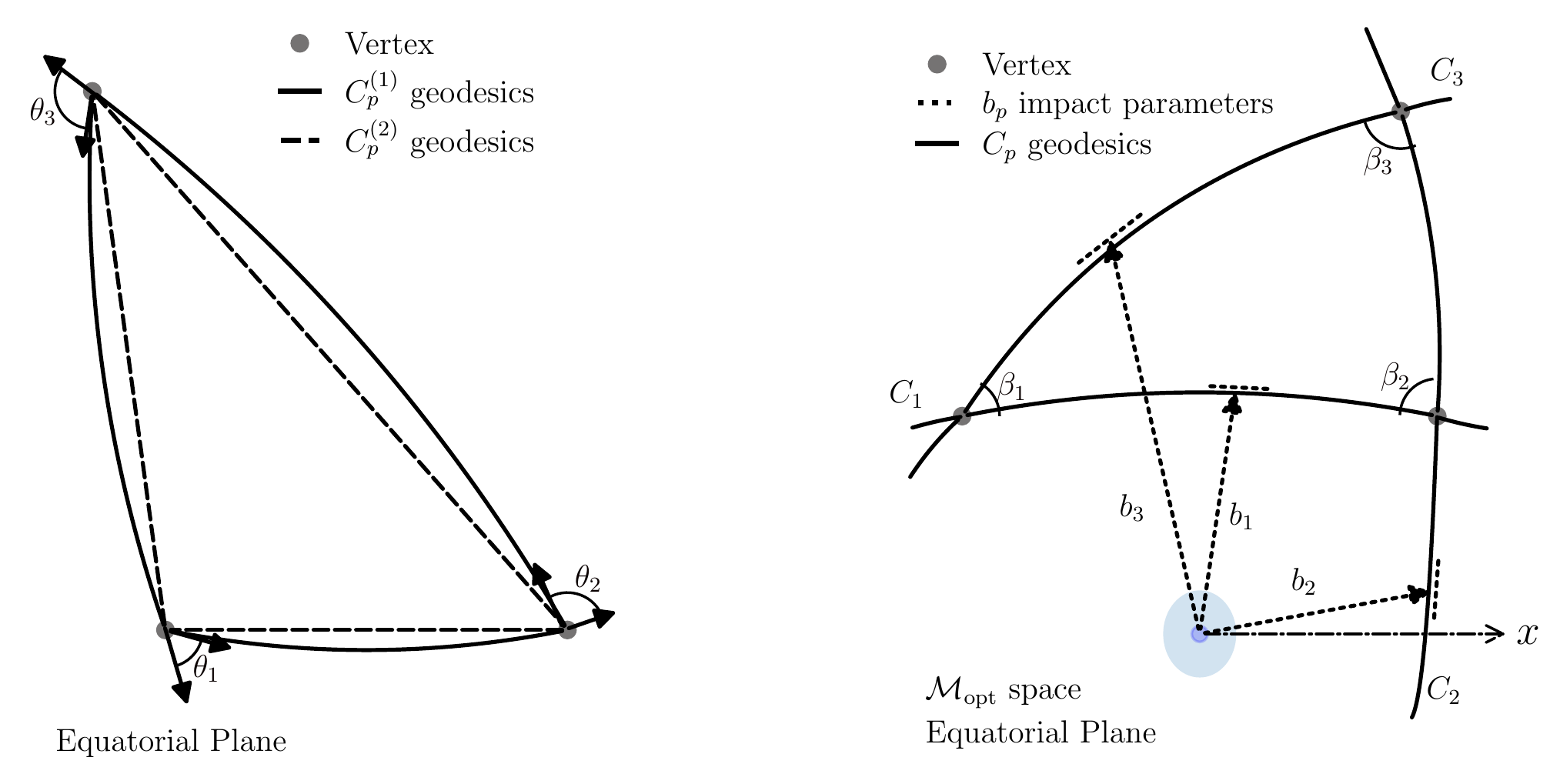}
	\caption{The left panel outlines the triangular configurations in different spacetimes. Solid lines represent the geodesics $C_{p}^{(I)}$ in the metric under study, while dashed lines represent geodesics in a background metric (Euclidean in this figure). From 
 the external angles $\theta_{p}$ one
 can obtain the internal angles $\beta_p=\pi-\theta_p$, which are used in
 the right panel, where we display a more accurate
 diagram of the setup proposed by H. Arakida~\cite{Arakida:2020xil}.}\label{Fig:Esq}
\end{figure*}

Exploiting the spherical symmetry, we impose without loss of generality that the photon orbits are in the equatorial plane $\theta=\pi/2$, reducing the optical metric to a $2$D-Riemannian space $\mathcal{M}_{\text{opt}}$. Now, we consider a region on $\mathcal{M}_{\text{opt}}$ consisting of an orientable triangular surface (represented as $\triangle$ from now on) bounded by three differentiable, parameterised curves $C_p$. From their respective geodesic curvatures,
\begin{equation}
\kappa_g(t)=\frac{1}{2\sqrt{\bar{g}_{rr}\bar{g}_{\phi\phi}}}\left(\frac{\partial \bar{g}_{\phi\phi}}{\partial r}\frac{d\phi}{dt}-\frac{\partial \bar{g}_{rr}}{\partial \phi}\frac{dr}{dt}\right)+\frac{d\Phi}{dt},\label{kappa}
\end{equation} 
one can arrive at the (local) Gauss-Bonnet theorem (for more details and proof, see section 4.5 in Ref.~\cite{Diff_GeomCarmo})\footnote{The original idea Ref.~\cite{Ishihara:2016vdc} consider a polygon $\Sigma^n$ bounded by $n$ smooth and piecewise regular curves $C_{p}$ ($p=1,2,...,n$). However, in a subsequent work Ref.~\cite{Takizawa:2020egm}, the authors put $n=3$.},
\begin{align}
    \int \int_{\triangle} Kd\sigma + \sum_{p=1}^{3}\int_{Cp}\kappa_{g}d\ell+\sum_{p=1}^{3}\theta_{p}=2\pi, \label{GB theorem}
\end{align}
where $\Phi$ in~(\ref{kappa}) is a differentiable function that measures the positive angle $\theta_{p}$ from the jump between the curves (see e.g., the left panel in fig.~\ref{Fig:Esq}), $d\ell = \sqrt{dt^2}$ is the line element along the boundary with sign chosen such that it is compatible with the orientation of the surface, $d\sigma=\sqrt{\abs{\bar{g}}}drd\phi$ is the areal element, and $K$ denotes the Gaussian curvature of $\triangle$. This curvature can be computed as (see e.g., eq.~(16) in Ref.~\cite{Werner:2012rc} or alternatively Theorem 13.25 in Ref.~\cite{abbena2006modern})
\begin{align}
    K&=\frac{R_{r\phi r\phi}}{\bar{g}}=\frac{R}{2},\label{Ec:GaussCurv}
\end{align}
where the geometrical tensors are calculated using the optical metric on the equatorial plane.

\subsection{Curvature in a region of space and the deflection angle}\label{Sec:II.2}

The postulate proposed by H. Arakida in Refs.~\cite{Arakida:2017hrm, Arakida:2020xil} provides an alternative approach for calculating the bending angle of light in static, spherically symmetric, and non-asymptotically flat spacetimes. The method involves computing the internal angles of polygons that encode the local spacetime curvature, and determining an angle $\alpha$ based on the difference between the sums of these internal angles. The angle $\alpha$ is referred to as the deflection angle in~\cite{Arakida:2020xil}. However, as we will explain in detail in the following subsections, it should be noted that it only coincides with the standard definition of a deflection angle under certain approximations. Therefore, we prefer to describe $\alpha$ as an \emph{angular difference}. This angular difference is computed by placing $\triangle$ in an assumed background spacetime and another one in the spacetime under study and taking the difference
\begin{align}
    \alpha :=\abs{\sum_{p=1}^{3}\left(\beta_{p}^{(1)}-\beta^{(2)}_{p}\right)}, \label{eq:TDefAng}
\end{align}
where $\beta^{(I)}_{p}$ are the internal angles of the triangles that are placed in the spacetimes $I = 1,2$. Notice that $\alpha$ is defined as a positive value. Following Ref.~\cite{Arakida:2020xil}, we assign the label $I=1$ to the spacetime under investigation, while $I=2$ denotes the background metric. For instance, if we study the deflection due to an object of a mass $M$, the background metric $I=2$ is interpreted as the respective massless metric. The right panel in figure~\ref{Fig:Esq} shows a schematic triangle setup.

The angles $\beta^{(I)}_{p}$ can be obtained by applying the Gauss-Bonnet theorem~(\ref{GB theorem}) to each of the surfaces $\triangle^{(I)}$. For this, we use that the curves $C^{(I)}_p$ are geodesics in their respective spaces $\mathcal{M}^{(I)}_{\text{opt}}$, which implies that $\kappa^{(I)}_g=0$\footnote{From Fermat's principle (in a spacetime) we have that along the light ray curve $\delta\int dt=0$~\cite{VPerlick_1990}. From this, one can deduce the expression for the light geodesic curvature $\kappa_{g}$ in a stationary spacetime (see e.g. eq.~(13) in Ref.~\cite{Ono:2017pie}). In particular, for a spherically symmetric spacetime, the trajectory of light is a geodesic (a unit-speed curve) in $\mathcal{M}_{\text{opt}}$ which means that $\kappa_g=0$~\cite{oprea2007differentia}. This is equivalent to taking $a^{i}=0$ in Ref.~\cite{Ono:2017pie}.}. Therefore their line integral in eq.~(\ref{GB theorem}) vanishes. Subtracting the 
results of the Gauss-Bonnet theorem for each $\triangle^{(I)}$, we get 
\begin{align}
    \int \int_{{}\triangle^{(1)}} K^{(1)} d\sigma-\int \int_{{}\triangle^{(2)}} K^{(2)} d\sigma =\sum_{p=1}^{3}\left(\theta^{(2)}_{p}-\theta^{(1)}_{p}\right)=\sum_{p=1}^{3}\left(\beta^{(1)}_{p}-\beta^{(2)}_{p}\right),\label{Eq:TotalDef}
\end{align}
where we used the relation $\theta^{(I)}_{p}=\pi-\beta^{(I)}_{p}$
for supplementary adjacent angles (see left panel fig.~\ref{Fig:Esq}). Taking the absolute value we arrive at the angular difference
\begin{align}
    \alpha &= \abs{\sum_{p=1}^{3}\left(\beta_{p}^{(1)}-\beta^{(2)}_{p}\right)}=\abs{\int \int_{{}\triangle^{(1)}} K^{(1)} d\sigma-\int \int_{{}\triangle^{(2)}} K^{(2)} d\sigma} \,.\label{Eq:TotalAngl}
\end{align}
The right-hand side of the first equality is called \textit{the angular formula}~\cite{Arakida:2020xil} and it is used in our numerical implementation. The right-hand side of the second equality is called
\textit{the integral formula}~\cite{Arakida:2020xil} and it is used in the next subsections to find  iterative analytical solutions. In both cases, we need to compute the respective null geodesics. 

From the null condition $ds^2 = 0$, we have that for metrics with the structure of eq.~(\ref{metric}) the orbit equation reduces to
\begin{align} 
    \left(\frac{dr}{d\phi}\right)^2+f(r) r^2=\frac{r^4 f(r)}{b^2 h(r)}, \label{Eq:Orbtray}
\end{align}
where $b:=L/E=[r^2/h(r)]d\phi/dt$ is the impact parameter defined in terms of the photon energy $E=h(r)dt/d\lambda$ and angular momentum $L=r^2d\phi/d\lambda$, with $\lambda$ the affine parameter along the light curve. The photon energy and angular momentum are conserved due to the Killing vector fields of the metric~(\ref{metric}). Also, using the conformal relation between the spacetime and optical metrics, one can prove that the null geodesics are the same in both metrics.

\subsection{The Kottler spacetime}\label{Sec:II.3}

In this section, we revisit the calculation of
the angular difference~(\ref{Eq:TotalAngl})
in the Kottler spacetime, where the metric functions in eq.~(\ref{metric}) are defined as
\begin{align} \label{}
    h(r)&=f(r), \qquad f(r)=1-\frac{2GM}{r}-\frac{\Lambda r^2}{3}.\label{Eq:KottlerMet}
\end{align}
This choice corresponds to a solution of the Einstein equations found by Kottler in $1918$\footnote{The solution~(\ref{Eq:KottlerMet}) was found independently by Kottler~\cite{Kottler1918} and Weyl~\cite{weyl1919statischen}, but it is commonly called Kottler solution.} and describes the effect of a cosmological constant $\Lambda$ on the spacetime around a spherically symmetric, static source of mass $M$. When $\Lambda=0$, the Schwarzschild metric is recovered, and when $\Lambda$ is positive (negative) the solution is also known as the Schwarzschild–de Sitter (Schwarzschild-anti-de Sitter) metric. In these two cases, the metric is non-asymptotically flat. Indeed, the Ricci scalar curvature of the Kottler spacetime is constant, $R=4\Lambda$. In such cases, the pedagogical methodology to compute the deflection angle is formally not viable, and the light source/observer has to be kept at a finite distance. A possibility to address this issue is to calculate the angular difference, eq.~(\ref{Eq:TotalAngl}). In this case, the background metric is eq.~(\ref{Eq:KottlerMet}) without a baryonic source, i.e. $M=0$.

Using the relations~(\ref{eq:optmetric}) together with the Kottler metric functions~(\ref{Eq:KottlerMet}) we compute the non-vanishing components of the optical metric on the equatorial plane,
\begin{equation}\label{eq:optmetricKotter}
    \bar{g}_{rr}=\left[1-r_{s}u\left(1+\frac{\lambda}{3 u^3}\right)\right]^{-2}, \ \ \ \bar{g}_{\phi\phi}=\frac{b^2}{u^2}\left[1-r_{s} u\left(1+\frac{\lambda}{3 u^3}\right)\right]^{-1},
\end{equation}	
where we introduced the dimensionless quantities 
\begin{align}\label{eq:DimLessVa}
    \begin{array}{lcl}
        u:=\dfrac{b}{r}, & \lambda:=\dfrac{\Lambda b^2}{r_s}=\dfrac{\bar{\Lambda}}{r_s},& r_{s}:=\dfrac{2GM}{b}.
    \end{array}
\end{align}
Notice that $r_s$ is a dimensionless version of the Schwarzschild radius, scaled with the impact parameter $b$. Similarly, $\tilde\Lambda$ is a dimensionless cosmological constant, and $\lambda$ is another dimensionless quantity that combines contributions from $\Lambda$ and $M$. This parameterization differs from those used in other works. For example, in Ref.~\cite{Arakida:2011ty}, an effective impact parameter is defined, where the influence of the cosmological constant $\Lambda$ is absorbed, resulting in a similar equation to the Schwarzschild case. An interesting discussion of its implications is presented in Ref.~\cite{Ishihara:2016vdc}. A different strategy is examined in Ref.~\cite{Takizawa:2020egm}, where the authors employ a double expansion using the constant terms $\Lambda$ and $r_g:=2GM$ as expansion parameters. One advantage of our approach is that we keep a single, dimensionless expansion parameter, $r_s$. 

From the Gaussian curvature given in eq.~(\ref{Ec:GaussCurv}), and using the areal element on the equatorial plane $d\sigma=\sqrt{\abs{\bar{g}}}drd\phi$ we compute the surface integrals in eq.~(\ref{Eq:TotalAngl}) as
\begin{align}
A^{(I)}=&\int_{\phi_{1}}^{\phi_{2}}\int_{u_{1}}^{u_{2}} du d\phi \left(\frac{r_s\lambda}{3 u^{3}}-r_s\left[\frac{r_s \lambda}{u^2}+\frac{3 r_s u}{4}-1\right]\right) \left[1-r_s u\left(1+\frac{\lambda}{3 u^3}\right)\right]^{-3/2},\nonumber\\
=&\int_{\phi_{1}}^{\phi_{2}}d\phi \,G(u) \bigg|_{u_1}^{u_2}, \label{Eq:IntSurf}
\end{align}
where
\begin{align}
    G(u):= -\left(1-\frac{3 r_s u}{2}\right)\left[1-r_s u\left(1+\frac{\bar{\Lambda}}{3u^3 r_s}\right)\right]^{-1/2}, \label{Eq:IntegrandSurf}
\end{align}
is the exact result of the integration over $u$,
and we used that $\lambda=\bar{\Lambda}/r_s$. It is important to note the following points. First, the quantity $u$ corresponds to the geodesics $C_p$ that bound $\triangle^{(I)}$ in each spacetime $I$, and it depends on the angular position $\phi$ as determined by the orbit equation~(\ref{Eq:Orbtray}). Second, we choose to calculate the (dimensionless) surface integral $A^{(I)}$ as the sum of the region under each side of the triangle, for instance, $A^{(1)}=A^{(1)}_{C_1}+A^{(1)}_{C_2}+A^{(1)}_{C_3}$, with the angular limits of integration placed in such a way that the area below $C_1$ cancels out. The curves $C_p$ are null geodesics described by the orbit equation~(\ref{Eq:Orbtray}), which for the Kottler spacetime~(\ref{Eq:KottlerMet}) in terms of the rescaled quantities~(\ref{eq:DimLessVa}) is reduced to
\begin{equation}
    \left(\frac{d u}{d\phi}\right)^2+u^2\left(1-r_s u\right)=1+\frac{r_s\lambda}{3}. \label{Eq:GeoK2}
\end{equation}
From eq.~(\ref{Eq:TotalAngl}), and using eqs.~(\ref{Eq:IntSurf},~\ref{Eq:GeoK2}), we can compute the angular difference $\alpha$. However, obtaining an exact solution is extremely difficult, or even impossible. Therefore, we resort to iterative and numerical methods. Iterative solutions valid in some regions of the parameter space are explored in this section, and a numerical approach is presented in section~\ref{Sec:III}.
\begin{figure*}[t]
	\centering	 
\includegraphics[width=11.cm]{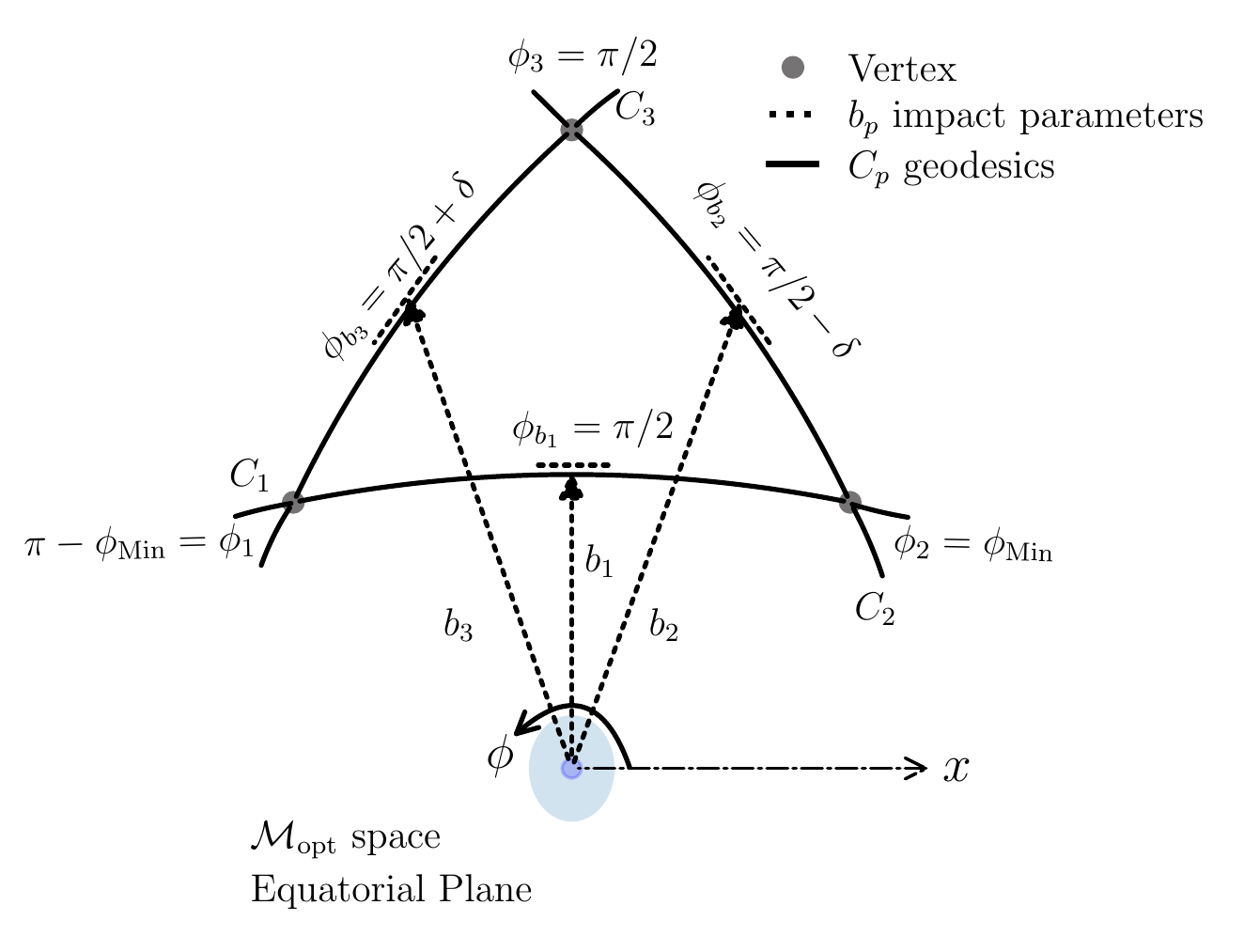}
	\caption{Triangular configuration used to compute the angular difference $\alpha$, eq.~(\ref{Eq:TotalAngl}). Note that for every $C_p$ geodesic, we have an impact parameter $b_p$ with an angular position $\phi_{b_{p}}$, where $p=1,2,3$.
 }\label{Fig:Esq2}
\end{figure*}

Let  us assume a configuration consisting of three geodesics that outline a triangular array placed symmetrically with respect to the vertical axis, as shown in fig.~\ref{Fig:Esq2}. Furthermore, we perform an expansion of $u$ in terms of a small dimensionless parameter $0 <r_s\ll 1$,\footnote{The expansion~(\ref{Eq:uExpans}) is valid for the weak deflection regime. We assume a light ray that is unaffected by $r_s$ or, equivalently, by $M$ and $\bar\Lambda$. The different mathematical procedures were checked using the software Wolfram Mathematica and our notebooks are publicly available in~\cite{Roque_On_the_radial_2023}.}
\begin{align}
    u^{(C_p, I)}(\phi) = u_0(\phi) +r_s u_1(\phi) +r_s^2 u_2(\phi)+\mathcal{O}(r_s^{3}),\label{Eq:uExpans}
\end{align}
where the superscript $(C_p, I)$ indicates the geodesic and the metric. Introducing the expansion~(\ref{Eq:uExpans}) into eq.~(\ref{Eq:GeoK2}), the equation is rewritten as a power series in $r_s$, with coefficients that depend on $u_0, u_1, u_2$ and their derivatives. Solving order by order for these coefficients we obtain the solutions for $u^{(C_p, I)}$. However, for the purpose of writing this solution, we discard terms of order higher than $\mathcal{O}(r_s^{2}, \bar{\Lambda}^2)$, obtaining\footnote{The iterative solution eq.~(\ref{Eq:uSerie}) obtained from our expansion~(\ref{Eq:uExpans}) coincides with the result reported in eq.~($18$) of Ref.~\cite{Takizawa:2020egm} to order $\mathcal{O}(r_s^{2}, \bar{\Lambda}^2)$, and with eq.~($6.5$) in Ref.~\cite{Arakida:2020xil} after expanding the effective impact parameter $B_{i}$.}
\begin{align}
    u^{(C_p, I)} = \xi_0\left(1+\frac{\bar{\Lambda}}{6}-\frac{\bar{\Lambda}^2}{72}\right)+\xi_1\left(1+\frac{\bar{\Lambda}}{3}\right)r_s+\xi_2\left(1+\frac{\bar{\Lambda}}{2}+\frac{\bar{\Lambda}^2}{24}\right)r_s^2+\mathcal{O}(r_s^{3}, \bar{\Lambda}^3),\label{Eq:uSerie}
\end{align}
with
\begin{subequations}
\begin{align}
    \xi_0&:=\sin(\phi-\delta_p),\\
    \xi_1&:=\frac{ 3+\cos(2\phi-2\delta_p)}{4},\\
    \xi_2&:=\frac{37\sin(\phi-\delta_p)-3\sin(3\phi-3\delta_p)+30\Phi\cos(\phi-\delta_p)}{64},
\end{align}
\end{subequations}
where we defined $\Phi:=\pi-2(\phi-\delta_p)$ and $\delta_p$ is the angular position at which $du^{(C_p, I)}/d\phi\big|_{\pi/2+\delta_p}=0$. From fig.~\ref{Fig:Esq2}, we have that for the null geodesics $C_p$ the angular position $\delta_p$ is
\begin{align}\label{eq:Cp_Delta}
    \begin{array}{lcl}
        C_1 \to \delta_1=0, & C_2 \to \delta_2=-\delta,& C_3 \to \delta_3=\delta,
    \end{array}
\end{align}
where $\delta$ is a constant parameter that quantifies the angular deviation with respect to the impact parameter $b_1$ corresponding to the null geodesic $C_1$. Notice that we fixed the angular coordinate of the impact parameters to $\phi_{b_p}=\pi/2+\delta_p$ (see fig.~\ref{Fig:Esq2}).

\setcounter{footnote}{0} 
Now, we evaluate the (iterative) geodesic solutions~(\ref{Eq:uSerie}) into eq.~(\ref{Eq:IntegrandSurf}) and express $G(u)$ as a second-order Taylor expansion around a flat region of spacetime (i.e., we expand around $r_s=0$ and $\bar{\Lambda}=0$). After integrating over the angular component in eq.~(\ref{Eq:IntSurf}) we arrive to,
\begin{align}
    A^{(I)}= \mathcal{A}^{(C_1, I)}\bigg|_{\phi_{1}}^{\phi_{2}}+\mathcal{A}^{(C_2, I)}\bigg|_{\phi_{2}}^{\phi_{3}}+\mathcal{A}^{(C_3, I)}\bigg|_{\phi_{3}}^{\phi_{1}},\label{Eq:AIterativSD}
\end{align}
where the limits of integration $\phi_i$ are given in fig.~\ref{Fig:Esq2}, and
\begin{align}
    \mathcal{A}^{(C_p, I)}= \Gamma_{0}^{(C_p)}+\frac{2GM}{b_p} \Gamma_{1}^{(C_p)}+\left(\frac{2GM}{b_p}\right)^2 \Gamma_{2}^{(C_p)}+\mathcal{O}\left(\frac{2GM}{b_p}\right)^{3},
\end{align}
where the expression for $\Gamma_{i}^{(C_p)}$ are reported in~\ref{Sec:Appendix}, eqs.~(\ref{GammaExp}). Notice that we used the re-scaling~(\ref{eq:DimLessVa}) to express $A^{(I)}$ in physical units ($\hbar=c=1$). The sub-index $p$ on the parameters $\delta$ and $b$ denotes that these correspond to the null geodesic $C_p$.

Finally, using eq.~(\ref{Eq:AIterativSD}), we have that for an array $\triangle$ like the one shown fig.~\ref{Fig:Esq2}, the angular difference (\ref{Eq:TotalAngl}) in a Schwarzschild–de Sitter spacetime ($\Lambda>0$) -- whose background is a de Sitter spacetime -- is 
\begin{align}
    \alpha&= \Bigg|4GM\bigg\lbrace \frac{\cos([\delta+\phi_2])+\sin(\delta)}{b_2}-\frac{\cos(\phi_2)}{b_1}+\frac{\Lambda}{24}\alpha_{11}+\frac{\Lambda^2}{144}\alpha_{12}\bigg\rbrace\nonumber\\
    &+\frac{G^2M^2}{4}\bigg\lbrace\frac{1}{b_1^2}\left[\sin(2\phi_2)-15\pi\bigg(1-\frac{b_1^2}{b_2^2}\left[1-\frac{2\phi_2}{\pi}\right]-\frac{2\phi_2}{\pi}\bigg)-\frac{2 b_1^2}{b_2^2}\cos(\phi_2)\sin(\phi_2+2\delta)\right]\nonumber\\
    &+\frac{\Lambda}{6}\alpha_{21}+\frac{\Lambda^2}{96}\alpha_{22}\bigg\rbrace\Bigg|, \label{Eq:AngSchDSitter}
\end{align}
where the expressions for $\alpha_{ij}$ are given in~\ref{Sec:Appendix}, eqs.~(\ref{AlphaKot}). At this point, it is important to remark on some aspects of eq.~(\ref{Eq:AngSchDSitter}):
\begin{itemize}
\item[i.] Every term in $\alpha$ -- including those involving $\Lambda$ -- is multiplied by some power of $GM$. Thus, when $M = 0$, we get $\alpha=0$. This is precisely what is expected since the background spacetime is taken as de Sitter.
\item[ii.] The result is in agreement with Ref.~\cite{Arakida:2020xil}, 
this is established by expanding their effective impact parameter in terms of $b$ and $\Lambda$.
\item[iii.] For $\Lambda=0$, and a triangular configuration where the impact parameter $b_2=b_3\gg b_1$ the angular difference $\alpha$ can be approximated as 
\begin{align}
    \alpha^{\infty} \approx \frac{4GM}{b_1}+\frac{15\pi G^2 M^2}{4 b_1^2}-\left(\frac{2GM}{b_1}+\frac{8G^2M^2}{b_1^2}\right)\phi_{2}^{2}+\mathcal{O}\left(\phi_{2}^3\right).\label{SchAng}
\end{align}
As can be noted, the first term on the right-hand side corresponds to the post-Newtonian result for the Schwarzschild metric (see for example~\cite{Misner:1973prb}, or section $7.1$ in~\cite{will_2018}, or $5$ in~\cite{Weinberg:1972kfs}), and the second term is the post-post-Newtonian correction~\cite{Epstein:1980dw}. The rest of the terms are the corrections due to the finite size of the triangles, which appear as second-order corrections in $\phi_2$ and always reduce the size of $\alpha^\infty$. The formula for the deflection angle of the Schwarzschild metric is recovered in the limit $\phi_2\to0$ -- which implies that $b_2=b_3\to\infty$. This means that in this limit the triangular configuration would behave as if the light source and the observer were in the flat region at spatial infinity. In section~\ref{Sec:III} this will be discussed in more detail.
\item[iv.] The coefficients $\alpha_{ij}$, presented in~\ref{Sec:Appendix}, contain  positive powers of $b_1, b_2$, however, their contribution to the angular difference is weighted by $\Lambda$, therefore unrealistic impact parameters in the order of $10^{26}$\,m would be needed for these terms to be comparable to the zero-order terms. On the other hand, in the term $GM\Lambda\alpha_{11}/6$, one can identify, for small $\phi_2$ values, the well-known $2Mb_1\Lambda/3$ contribution characteristic of the Kottler spacetime~\cite{Rindler:2007zz, Ishak:2007ea}.

\end{itemize}
\subsection{A Horndeski gravity spacetime}\label{Sec:II.4}

Hordenski gravity~\cite{Horndeski:1974wa, Deffayet:horndeski, Deffayet:horndeski2} is the most general local and Lorentz invariant scalar-tensor theory with second-order equations of motion with an extra degree of freedom (apart from the usual spin two metric field). A characteristic of this theory is the presence of combinations of non-canonical kinetic terms and non-minimal couplings between curvature  tensors and the scalar field. This gives the theory a rich phenomenology in the context of compact objects, see e.g., the Refs.~\cite{Chagoya:2020bqz, Afrin:2021wlj, Barranco:2021auj, Roque:2021lvr, Atamurotov:2022slw}.

In this paper we focus on a non-asymptotically flat spacetime whose metric does not have the same structure as Kottler. This spacetime is a solution to the shift symmetric ($\varphi\to\varphi+c$ ) subset of  Horndeski gravity studied in Ref.~\cite{Babichev:2016rlq}, 
\begin{align}
  S &= \frac{M_{\textrm{Pl}}^2}{2}\int d^4 x \sqrt{-g}\bigg(R-2\Lambda-\frac{1}{2} \nabla_{\mu}\varphi \nabla^{\mu}\varphi+ \beta G^{\mu \nu} \nabla_{\mu}\varphi\nabla_{\nu}\varphi \bigg), \label{Eq:HLagrangian}
\end{align}
where the massless scalar field $\varphi$ is dimensionless and $\beta$ is the Horndeski coupling constant and has units of length squared. The line element for the solution that we use (Class II in~\cite{Babichev:2016rlq}) corresponds to eq.~(\ref{metric}) with 
\begin{align} 
    h(r)&=1-\frac{2GM}{r},\qquad f(r)=\left(1-\frac{2GM}{r}\right)\left(1-r^{2}\Lambda\right),\label{Eq:HorndMetric}
\end{align}
where $M$ corresponds to the mass of the black hole, $G$ represents the gravitational constant, and $\Lambda$ is a constant. Furthermore, we establish the relationship $2\beta\Lambda=-1$ for the parameter $\beta$. In addition a time-dependent Ansatz for the scalar field $\phi(t,r):= qt+\psi(r)$ is used. The radial part of the scalar is determined by the differential equation
\begin{align}
    \frac{d\psi}{dr}&=\pm\frac{q}{h(r)}\sqrt{\frac{2GM}{r\left(1-r^2\Lambda\right)}},\nonumber
\end{align}
where $q$ is an integration constant. It is easy to check that for the solution~(\ref{Eq:HorndMetric}), the Ricci scalar reduces to
\begin{align}
R=6\Lambda-\frac{6GM\Lambda}{r}\,,\nonumber
\end{align}
which means that for large $r$ it behaves as $R\sim4\Lambda_{\text{eff}}$, where $\Lambda_{\text{eff}}:=3\Lambda/2$. In other words, for a large radius, Horndeski's contribution drives the curvature scalar $R$ to exhibit a behaviour equivalent to that of a Kottler spacetime. However, for short or medium radii, we anticipate the presence of discrepancies arising from the alternative model. It is also important to note that, unlike the Kottler solution~(\ref{Eq:KottlerMet}) where $\Lambda$ is constrained to the observational value of the cosmological constant, $\Lambda\approx 1.1\times 10^{-52} \text{m}^{-2}$~\cite{Planck:2018vyg}, for the Horndeski solution~(\ref{Eq:HorndMetric}), $\Lambda$ does not necessarily play the role of the observed cosmological constant, since this solution belongs to a modified gravity model where the cosmological dynamics might differ from $\Lambda$-CDM. For this reason, when working with the solution~(\ref{Eq:HorndMetric}) we allow for different values of $\Lambda$ and we study their effect on the angular difference. Also, when required to avoid confusion, we will refer to the constant $\Lambda$ that appears in the Lagrangian as ``bare cosmological constant'', and to the cosmological constant fitted observationally in the context of $\Lambda$CDM as ``observed cosmological constant''.

In this subsection, we apply to the solution~(\ref{Eq:HorndMetric}) the methodology that was used in the previous subsection for the Kottler spacetime. Our motivations are twofold: on the one hand, to further investigate Arakida's proposal in spacetimes that go beyond asymptotically flat scenarios and beyond GR\footnote{In~\cite{Javed:2020pyz}, the Gauss-Bonnet theorem was used to compute the bending angle of light in the weak field limit of an asymptotically flat black hole solution to Horndeski gravity.}, and on the other hand, to identify differences with respect to the bending angle predicted in other spacetimes that could serve as proposals for observational tests.

From the relations~(\ref{eq:optmetric}) we have that the non-vanishing components of the {optical metric} on the equatorial plane to the solution~(\ref{Eq:HorndMetric}) are
\begin{align}\label{eq:optmetricHorndeski}
    \begin{array}{lcl}
    \bar{g}_{rr}=\left(1-r_{s}u\right)^{-2}\left(1-\frac{r_{s}\lambda}{u^2}\right)^{-1}, &  &\bar{g}_{\phi\phi}=\frac{b^2}{u^2}\left(1-r_{s}u\right)^{-1},
    \end{array}
\end{align}	
where we used the dimensionless quantities defined in~(\ref{eq:DimLessVa}). Using the optical metric we compute the surface integrals, eq.~(\ref{Eq:TotalAngl}), obtaining 
\begin{align}
A^{(I)}=&\int_{\phi_{1}}^{\phi_{2}}\int_{u_{1}}^{u_{2}} du d\phi  \left[\left(1+\frac{3r_s \lambda}{2u^2}-\frac{3r_s^2\lambda}{4u}-\frac{3 r_s u}{4}\right)r_s-\frac{\lambda r_s}{u^{3}}\right]\left[\left(1-\frac{\lambda r_s}{u^2}\right)\left(1-u r_s\right)^3\right]^{-1/2},\nonumber\\
=&\int_{\phi_{1}}^{\phi_{2}}d\phi \,G(u) \bigg|_{u_1}^{u_2}, \label{Eq:IntSurfHornd}
\end{align}
where $G(u)$ is the exact result of the integration over $u$,
\begin{align}
    G(u):= -\left(1-\frac{3r_{s}u}{2}\right)\left(1-\frac{\bar{\Lambda}}{u^2}\right)^{1/2}\left(1-r_{s}u\right)^{-1/2}. \label{Eq:IntegrandSurfHornd}
\end{align}
Since the Horndeski model~(\ref{Eq:HLagrangian}) satisfies the weak equivalence principle, i.e., test particles move on geodesics, the orbit equation~(\ref{Eq:Orbtray}) is valid, and for the solution~(\ref{Eq:HorndMetric}) it reduces to
\begin{align}
    \left(\frac{d u}{d\phi}\right)^2+r_s \lambda\left(\frac{1}{u^2}+r_s u\right)+u^2\left(1-r_s u\right)=1+r_s\lambda.\label{Eq:GeoHord}
\end{align}
Using the expansion~(\ref{Eq:uExpans}), we obtain the iterative solution reported in Appendix A eq.~(\ref{AEq:uSerieHord}), whose first terms are:
\begin{align}
u^{(C_p, I)} &= \sin(\phi-\delta_p)\left(1+\frac{\cot^2{(\phi-\delta_p)}}{2}\bar{\Lambda}-\mathcal{O}(\bar{\Lambda}^2)\right)+\frac{3}{4}\bigg\lbrace 1+\frac{\cos{(2\phi-2\delta_p)}}{3}\nonumber\\
&-\cot^2{(\phi-\delta_p)}\left[1-\frac{\cos{(2\phi-2\delta_p)}}{3}\right]\bar{\Lambda}+\mathcal{O}(\bar{\Lambda}^2)\bigg\rbrace r_s+\mathcal{O}(r_s^{2}, \bar{\Lambda}^2).\label{Eq:uSerieHord}
\end{align}
Inserting eq.~(\ref{AEq:uSerieHord}) into eq.~(\ref{Eq:IntSurfHornd}) and employing the same methodology as used for the Kottler solution we arrive at
\begin{align}
    A^{(I)}= \mathcal{A}^{(C_1, I)}\bigg|_{\phi_{1}}^{\phi_{2}}+\mathcal{A}^{(C_2, I)}\bigg|_{\phi_{2}}^{\phi_{3}}+\mathcal{A}^{(C_3, I)}\bigg|_{\phi_{3}}^{\phi_{1}},\label{Eq:AIterativ}
\end{align}
where we used the same triangular configuration shown in fig.~\ref{Fig:Esq2} and the terms $\mathcal{A}^{(C_p, I)}$ are given by
\begin{align}
    \mathcal{A}^{(C_p, I)}= \Gamma_{0}^{(C_p)}+\frac{2GM}{b_p} \Gamma_{1}^{(C_p)}+\left(\frac{2GM}{b_p}\right)^2 \Gamma_{2}^{(C_p)}+\mathcal{O}\left(\frac{2GM}{b_p}\right)^{3},
\end{align}
with the $\Gamma_{i}^{(Cp)}$ expression given in the Appendix eq.~(\ref{HorCp}).

Finally, using the previous results and eq.~(\ref{Eq:TotalAngl}) we find the angular difference
\begin{align}
    \alpha&= \Bigg|4GM\bigg\lbrace \frac{\cos([\delta+\phi_2])+\sin(\delta)}{b_2}-\frac{\cos(\phi_2)}{b_1}+\frac{\Lambda}{8}\alpha^{H}_{11}+\frac{\Lambda^2}{128}\alpha^{H}_{12}\bigg\rbrace\nonumber\\
    &+\frac{G^2M^2}{4}\bigg\lbrace\frac{1}{b_1^2}\left[\sin(2\phi_2)-15\pi\bigg(1-\frac{b_1^2}{b_3^2}\left[1-\frac{2\phi_2}{\pi}\right]-\frac{2\phi_2}{\pi}\bigg)-\frac{2 b_1^2}{b_2^2}\cos(\phi_2)\sin(\phi_2+2\delta)\right]\nonumber\\
    &+\frac{\Lambda}{2}\alpha^{H}_{21}+\frac{3\Lambda^2}{4}\alpha^{H}_{22}\bigg\rbrace\Bigg|.\label{Eq:AngHorndeski}
\end{align}
Notice that, at zero order in $\Lambda$, this is the same as that reported for the Kottler solution~(\ref{Eq:AngSchDSitter}). This implies that the Schwarzschild deflection angle~(\ref{SchAng}) is recovered in the limit $\phi_2\to0$ when $\Lambda=0$. The terms $\alpha_{ij}^{H}$ are reported in eq.~(\ref{AlphaKot}) in Appendix A.

\section{Numerical Analysis}\label{Sec:III}
In the previous section, we discussed iterative solutions of the integral formula for the angular difference, as given in eq.~(\ref{Eq:TotalAngl}). Specifically, we explored these solutions for two non-asymptotic flat spacetimes: the Kottler metric, eq. (\ref{Eq:KottlerMet}), and a solution derived from Horndeski gravity, eq.~(\ref{Eq:HorndMetric}). We considered a triangular configuration, as depicted in fig.~\ref{Fig:Esq2}, and obtained the corresponding iterative solutions for each metric, namely (\ref{Eq:AngSchDSitter}) and (\ref{Eq:AngHorndeski}). In this section, we extend our analysis by computing the deflection angle (\ref{Eq:TotalAngl}) through the evaluation of the internal angles $\beta^{(I)}_{p}$ of the triangular configuration.

\subsection{Implementation} \label{Sec:III.1}
To perform the numerical
analysis we use triangles bounded by the geodesics $C_p$, separately in the spacetime under study and in the background spacetime. These geodesics
are determined by the orbit 
equation~(\ref{Eq:Orbtray}),
and the angular positions $\phi_{b_{p}}$ of their respective impact parameters $b_p$ are 
\begin{align}
    \phi_{b_{1}}=\pi/2,\quad \phi_{{b_{2}}}=\pi/2-\delta,\quad \phi_{{b_{3}}}=\pi/2+\delta, \nonumber
\end{align}
as shown in fig.~\ref{Fig:Esq2}. For the geodesic $C_1$ we use a Neumann boundary condition, $du/d\phi=0$ at $\phi=\pi/2$, while for the rest of the geodesics we use Dirichlet boundary conditions in such a way that the bottom vertex of $C_2$ ($C_3$) coincides with the right (left) vertex of $C_1$ at $\phi_2=\phi_{\text{Min}}$ ($\phi_3=\phi_{\text{Max}}$), where $[\phi_{\text{Min}}, \phi_{\text{Max}}]$ is the angular integration domain. The numerical integration of eq.~(\ref{Eq:Orbtray}) with the boundary conditions described above is performed using an adaptive explicit fifth-order Runge-Kutta routine, with errors estimated using a fourth-order routine~\cite{2020SciPy-NMeth, DORMAND198019, Lawrence1986SomePR}.

Once the triangles are constructed, we compute their internal angles $\beta_p$ using the relations\footnote{ The relations~(\ref{EqAngRel}) are the same that the relations eqs.~(4.9-4.12) presented in Ref.~\cite{Arakida:2020xil}. The discrepancy in the definitions for $\beta_3$ is a consequence of that in our numerical integration for the geodesic $C_3$ we perform it in the clockwise angular direction.}
\begin{subequations}\label{EqAngRel}
\begin{align}
\beta_1&:=\psi_{\phi_{\text{Max}}}^{(3)}-\psi_{\phi_{\text{Max}}}^{(1)},\\
\beta_2&:=\psi_{\phi_{\text{Min}}}^{(1)}-\psi_{\phi_{\text{Min}}}^{(2)},\\
\beta_3&:=\psi_{\pi/2}^{(2)}-\psi_{\pi/2}^{(3)},
\end{align}
\end{subequations}
where $\beta_p$ are the vertices of the triangle (fig.~\ref{Fig:Esq2}) and
$\psi_a^{(p)}$ indicates an angle computed for the geodesic $p$ at $\phi=a$ by means
of the tangent formula~\cite{Perlick:2021aok}
\begin{align}
\tan \left(\psi_{a}^{(p)}\right)&=\sqrt{\frac{\bar{g}_{\phi\phi}(r_p)}{\bar{g}_{rr}(r_p)}}\frac{d\phi}{dr_p}\bigg|_{\phi=a}. \label{EqTang}
\end{align}
For example, $\psi_{\phi_{\text{Max}}}^{(3)}$ is computed by evaluating eq.~(\ref{EqTang}) with the numerical solution for the geodesic $C_3$ and computing the inverse tangent function at $\phi=\phi_{\text{Max}}$. The radial derivative is calculated using second-order accurate central differences~\cite{Alfio2007}.

Finally, using the internal angles $\beta_p$ and the angular formula~(\ref{Eq:TotalAngl}), we compute the angular difference $\alpha$ for the spacetime solution under study.  Our code is publicly available in~\cite{Roque_On_the_radial_2023}.

\subsection{Kottler spacetime}\label{Sec:III.2}
\begin{figure*}[t]
	\centering	
    \includegraphics[width=15.5cm]{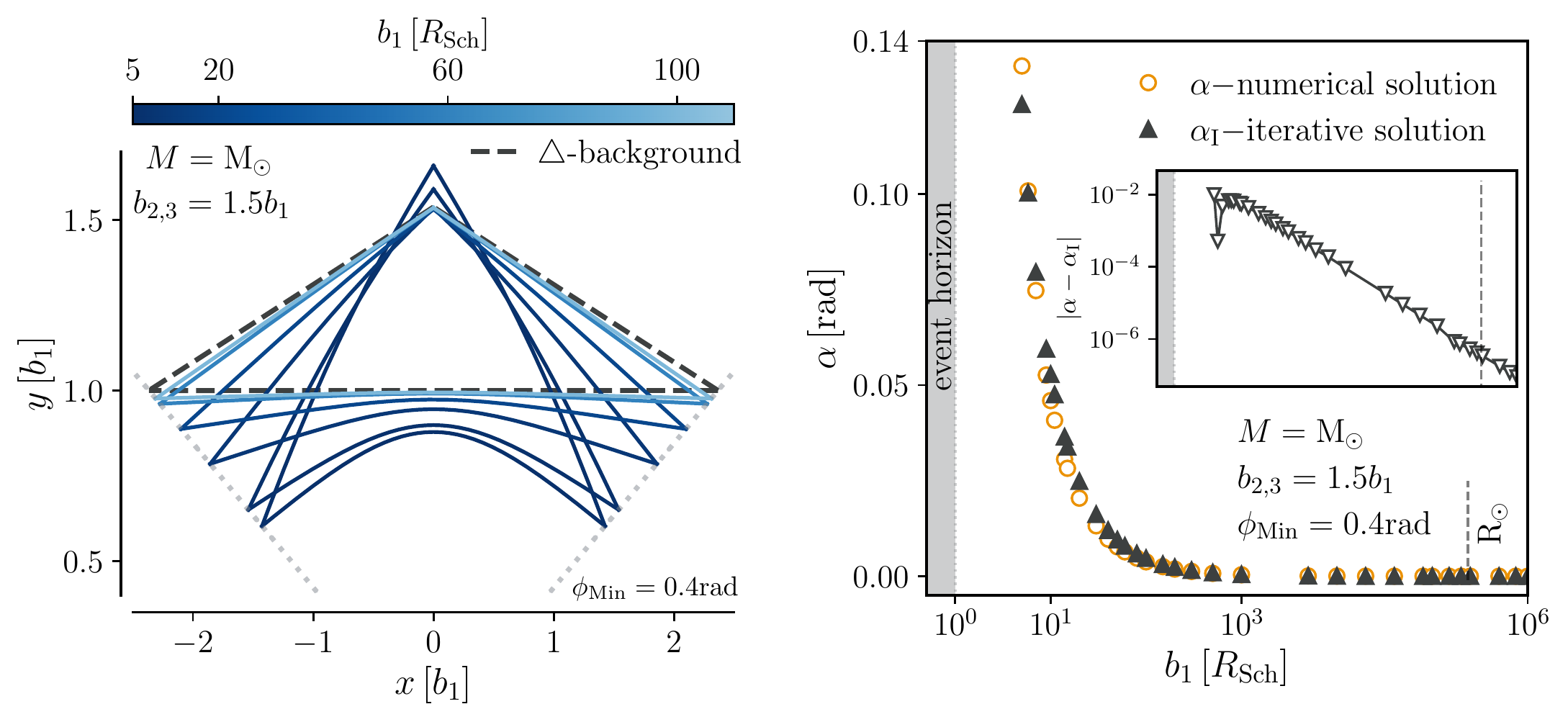}
	\caption{Influence of the impact parameter $b_1$. The left panel shows the change of a triangular configuration embedded in the Kottler spacetime (solid lines), in coordinates rescaled with $b_1$. The colour bar indicates the value of $b_1$, which determines how close the triangle is from the gravitational source. In these coordinates, the background $\triangle^{(2)}$ does not change. The right panel shows the change in the numerically computed angular difference $\alpha$ and using the iterative solution $\alpha_{I}$~(\ref{Eq:AngSchDSitter}). The insert exhibits their absolute difference. Observe that for $b_1\gtrsim 10 R_{\text{Sch}}$ both are in agreement. More details are discussed in the main text.} \label{Fig:NSchDe}
\end{figure*}

Following the steps described above, we construct a set of triangular configurations in two plausible scenarios, using the dimensionless variables given by eq.~(\ref{eq:DimLessVa}) for the Kottler solution~(\ref{Eq:KottlerMet}). First, we assume a gravitational source with a solar mass $M_1=\mathrm{M}_\odot$, and later we consider a source with a mass suitable for a galaxy, $M_2=10^{12} \mathrm{M}_\odot$. Based on our numerical results, we analyse the effect that the parameters $b_1$ and $b_2=b_3$ have on the sum of internal angles as well as potential signatures of the cosmological constant, which represents the effect of a non-asymptotically flat spacetime. Furthermore, we examine the validity of the iterative solution~(\ref{Eq:AngSchDSitter}). 

We start by analysing the effect of $b_1$ on the internal angles. It is important to note that this impact parameter determines the proximity of the triangular configuration to the gravitational source $M$. In order to gain some intuition, in the left panel of fig.~\ref{Fig:NSchDe} we plot, in Cartesian coordinates, triangular configurations for several values of $b_1$ in the interval $[5R_{\text{Sch}}, 10^6 R_{\text{Sch}}]$, where $R_{\text{Sch}}$ denotes the Schwarzschild's radius. In these configurations, we maintain fixed values for the parameters $b_2/b_1 = b_3/b_1 = 1.5$ and $\phi_{\text{Min}} = 0.4$~rad. The background configuration corresponds to the spacetime obtained by setting $M=0$ in the Kottler metric. We appreciate that small values of $b_1$ lead to elliptic triangles,  while large values of $b_1$ lead to configurations that approximate the background triangles (dashed lines). This behaviour is expected because as the impact parameter $b_1$ gets larger, the contribution of the central mass to the gravitational potential in the region where the triangle is located gets smaller. Moreover, it is important to observe that as we move farther away from the source, the impact parameter $b_1$ and the distance of closest approach become nearly identical.

Now we analyse the validity of the iterative solution~(\ref{Eq:AngSchDSitter}). The right panel of fig.~\ref{Fig:NSchDe} shows the angle $\alpha$ as a function of $b_1$. The orange circles represent the values obtained using a numerical implementation of the angular formula (\ref{Eq:TotalAngl}), while the black triangles correspond to the results from the iterative solution (\ref{Eq:AngSchDSitter}). In agreement with the left panel, the angular difference $\alpha$ decreases to zero as $b_1$ increases. In addition, for $b_{1}=\mathrm{R}_\odot$, $\alpha$ is of the same order of magnitude as the standard deflection angle~\cite{Misner:1973prb, will_2018, Weinberg:1972kfs} computed for a light ray coming from infinity. The inset in the right panel of fig.~\ref{Fig:NSchDe} displays the relative difference between the numerical and iterative calculations of $\alpha$. It allows the reader to corroborate that the iterative solution~(\ref{Eq:AngSchDSitter}) is a good approximation for medium and large values of $b_1$. However, higher differences appear when $b_1\lesssim 10 R_{\text{Sch}}$, where, by construction, the iterative solution is not a good approximation.
An alternative to address this regime is explored in Ref.~\cite{Chagoya:2020bqz}. At this point, it is important to remark that fig.~\ref{Fig:NSchDe} should exhibit a similar behaviour if we consider a different massive source, such as an object with  $M=10^{12}\mathrm{M}_\odot$. This similarity arises because the numerical construction of the triangles employs dimensionless variables~(\ref{eq:DimLessVa}), implying that a change in mass merely re-scales the sides of the triangles.

\begin{figure*}[t]
	\centering	
    \includegraphics[width=15.5cm]{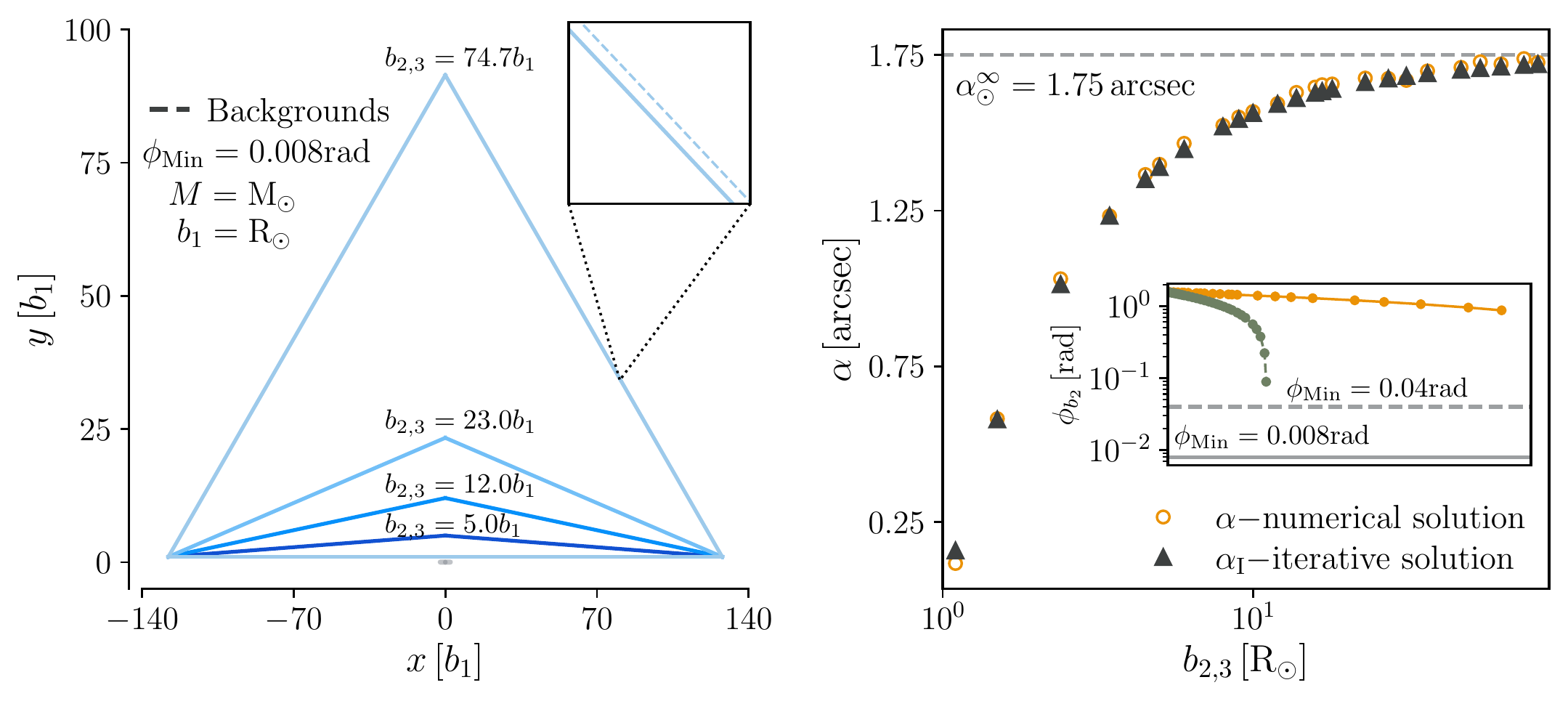}
	\caption{Influence of the impact parameters $b_2, b_3$ (abbreviated as $b_{2,3}$). The left panel exhibits the structural change of the triangles embedded in the background (dashed lines) and Kottler metrics (solid lines), around a source with $M=M_\odot$. Note that both are similar, this happens because we are far from the gravitational source (see fig.~\ref{Fig:NSchDe}). The right panel shows the angular difference computed numerically $\alpha$ and using the iterative solution $\alpha_{I}$~(\ref{Eq:AngSchDSitter}). For large values of $b_{2,3}$ we recover the observational deflection angle $\alpha^{\infty}=1.75$ arcsec. The inset shows the change in the angular position $\phi_{b_2}$ of the impact parameter $b_2$ when its value is increased. The dashed (solid) green (orange) line corresponds to $\phi_{\text{Min}}=0.04$rad ($\phi_{\text{Min}}=0.008$rad).  More details are discussed in the main text.}\label{Fig:NSchDe2}
\end{figure*}

Another limit of interest is $b_2=b_3\to\infty$ with small values of $\phi_2=\phi_{\text{Min}}$. As explained in subsection~\ref{Sec:II.3}, in this limit the iterative solution~(\ref{Eq:AngSchDSitter}) of the angular difference $\alpha$~(\ref{Eq:TotalAngl}) reproduces the Schwarzschild deflection angle $\alpha^{\infty}$ given in eq.~(\ref{SchAng}). This triangular configuration is highly unrealistic. However, from the results displayed in fig.~\ref{Fig:NSchDe2}, we see that $\alpha^{\infty}$ is already recovered for large but finite values of $b_2=b_3$. In the left panel of fig.~\ref{Fig:NSchDe2} we present triangular configurations for $b_1=\mathrm{R}_\odot$, $\phi_{\text{Min}}=0.008$ rad, and several values of $b_2=b_3$. Here, we consider a gravitational source with a mass of $M=\mathrm{M}_\odot$. It is worth noting that we are working within a regime where the iterative solution provides a good approximation (refer to fig.~\ref{Fig:NSchDe}). In the right panel of fig.~\ref{Fig:NSchDe2} we present the angular difference $\alpha$. We confirm that the numerical and iterative results are consistent. Furthermore, we observe that for $b_2=b_3\approx 80 \mathrm{R}_\odot$ the deflection angle $\alpha^{\infty}$ is recovered. Indeed, it is worth noticing that the impact parameters $b_2$ and $b_3$ cannot be arbitrarily large, since they cannot exceed the distance from the source to the lower vertices of the triangular configuration. This is confirmed in the inset of fig.~\ref{Fig:NSchDe2}, where we show that the angular position $\phi_{b_2}$ corresponding to the impact parameter $b_2$ approaches $\phi_{\text{Min}}=0.008$ rad as $b_2$ is increased, which means that the closest approach to the source tends to the bottom vertex. The same can be corroborated for $\phi_{b_3}$. In fact, for values of $\phi_{\text{Min}}\gtrsim 0.04$ rad, we do not recover the  deflection angle $\alpha^{\infty}$ asymptotically, on the contrary, an abrupt change in the slope appears as a consequence of reaching the vertices, see e.g., inset in fig.~\ref{Fig:NSchDe2}.

\renewcommand{\tabcolsep}{9pt}
\renewcommand {\arraystretch}{1.4}
\begin{table*}[t]
	\centering
	\begin{tabular}{|l@{\hskip .08 in} c@{\hskip .2 in} c@{\hskip .15 in} c@{\hskip .15 in} c@{\hskip .15 in} c@{\hskip .1 in}|}
 \hline
   \multicolumn{6}{|c|}{$\phi_{\mathrm{Min}}=0.008$ rad}\\[0.1cm] \hline
		Source &  $b_{1}$ & $b_{2,3}$ & $\alpha$& $\alpha_{\Lambda}^{\mathcal{O}(\Lambda)}$ & $\alpha_{\Lambda}^{\mathcal{O}(\Lambda^2)}$ \\[-0.2cm]
         $[\mathrm{M}_\odot]$& $[\mathrm{R}_\odot]$ & $[b_{1}]$ & [arcsec] & [arcsec] & [arcsec] \\[0.1cm]
		 \hline
		 $1.0$ & $1.0$ & $100$ & $1.7259$ &$2.3717\times10^{-31}$ & $5.0147\times10^{-62}$\\[0.1cm]
         $10$ & $100$ & $100$ & $0.172587$ & $2.3729 \times10^{-28}$ & $5.020\times10^{-55}$\\[0.1cm]
         $100$ & $300$ & $100$ & $0.5753$ & $ 7.1179\times10^{-27}$ & $1.355\times10^{-52}$\\[0.1cm]
        $4.0\times10^6$ (Sgr A$^{*}$~\cite{SagittariusA}) & $10^{7}$ & $100$ & $0.6904$ & $ 9.4901\times10^{-18}$ & $2.0072\times10^{-34}$\\[0.1cm]
         $6.5\times 10^{9}$ (M87~\cite{EventHorizonTelescope:2019ggy}) & $10^{10}$ & $100$ & $1.1218$ & $1.5419\times10^{-11}$ & $3.261\times10^{-22}$\\[0.1cm]
		 $10^{12}$ & $7.0\times 10^{11}$ & $100$ & $2.4024$ & $1.7013\times10^{-7}$ & $1.8584\times10^{-14}$\\[0.1cm]
		\hline
	\end{tabular}
	\caption{Contributions of $\Lambda$ at first $\alpha_{\Lambda}^{\mathcal{O}(\Lambda)}$ and second $\alpha_{\Lambda}^{\mathcal{O}(\Lambda^2)}$ order in the iterative solution~(\ref{Eq:AngSchDSitter}) for a set of possible massive sources. For comparison, we set the value for $b_2=b_3=100b_1$ and choose a reasonable value of $b_1$ such that it can represent some possible astrophysical scenarios.
	\label{tabla_1}}
\end{table*}

Finally, our results suggest that for a gravitational source of mass $M=\mathrm{M}_\odot$, the contribution of the cosmological constant $\Lambda$ to the angular difference $\alpha$ is unlikely to be observed. In table~\ref{tabla_1} we show this contribution at first and second order in $\Lambda$, as extracted from the iterative solution~(\ref{Eq:AngSchDSitter}), and we see that for a system with $M=\mathrm{M}_\odot$, $\Lambda$ affects the estimation of $\alpha$ at order $10^{-31}$ arcsec. However, if we take a source with mass $M\sim 10^{12} \mathrm{M}_\odot$, such as a galaxy, the $\Lambda$ contribution becomes relevant for the angle $\alpha$ at order $10^{-7}$ arcsec, which is in the range of sensitivity of future missions, e.g., LATOR~\cite{Turyshev:2003wt}. In other words, our results suggest that we can use gravitational lensing of galaxies to detect the contribution of $\Lambda$ to the deflection angle $\alpha$. This agrees with the ideas presented in Ref.~\cite{Arakida:2020xil}, with the difference that our estimate does not require a relation between $b_2, b_3$ and the de Sitter horizon. 

It is important to emphasise that the angular difference $\alpha$ is not the same physical quantity as the total bending angle $\alpha_{T}$. Although it is true that when $b_2=b_3$ are large enough, the Schwarzschild deflection angle is recovered, $\alpha^{\infty}$ does not contain the contribution of $\Lambda$. As first proved by Rindler and Ishak~\cite{Rindler:2007zz}, and later by others~\cite{Ishak:2007ea, Sereno:2007rm, Sereno:2008kk}, a positive cosmological constant reduces the effect on the total bending angle, i.e., the $\Lambda$ contribution tends to weaken the gravitational attraction, which contributes negatively to $\alpha_T$. In contrast, the $\Lambda$ contribution to the angular difference $\alpha$ is positive as seen in table~\ref{tabla_1}.

\subsection{Horndeski results}\label{Sec:III.3}

In the following, we present our main results concerning the study of the angular difference for Horndeski's solution~(\ref{Eq:GeoHord}).
We construct a set of triangular configurations in plausible scenarios and analyse the influence of the impact parameters and of the cosmological constant on the measured angular difference. 
As in the Kottler case, our results indicate that when the triangular configuration is far from the gravitational source (i.e. for large $b_1$), it approximates the background triangle, and in the limit $b_2=b_3\to\infty$ with a small value of $\phi_{\text{Min}}$ the angular difference $\alpha$ reproduces the Schwarzschild deflection angle $\alpha^{\infty}$. Let us expand on these points.

\begin{figure*}[t]
	\centering	
    \includegraphics[width=15.5cm]{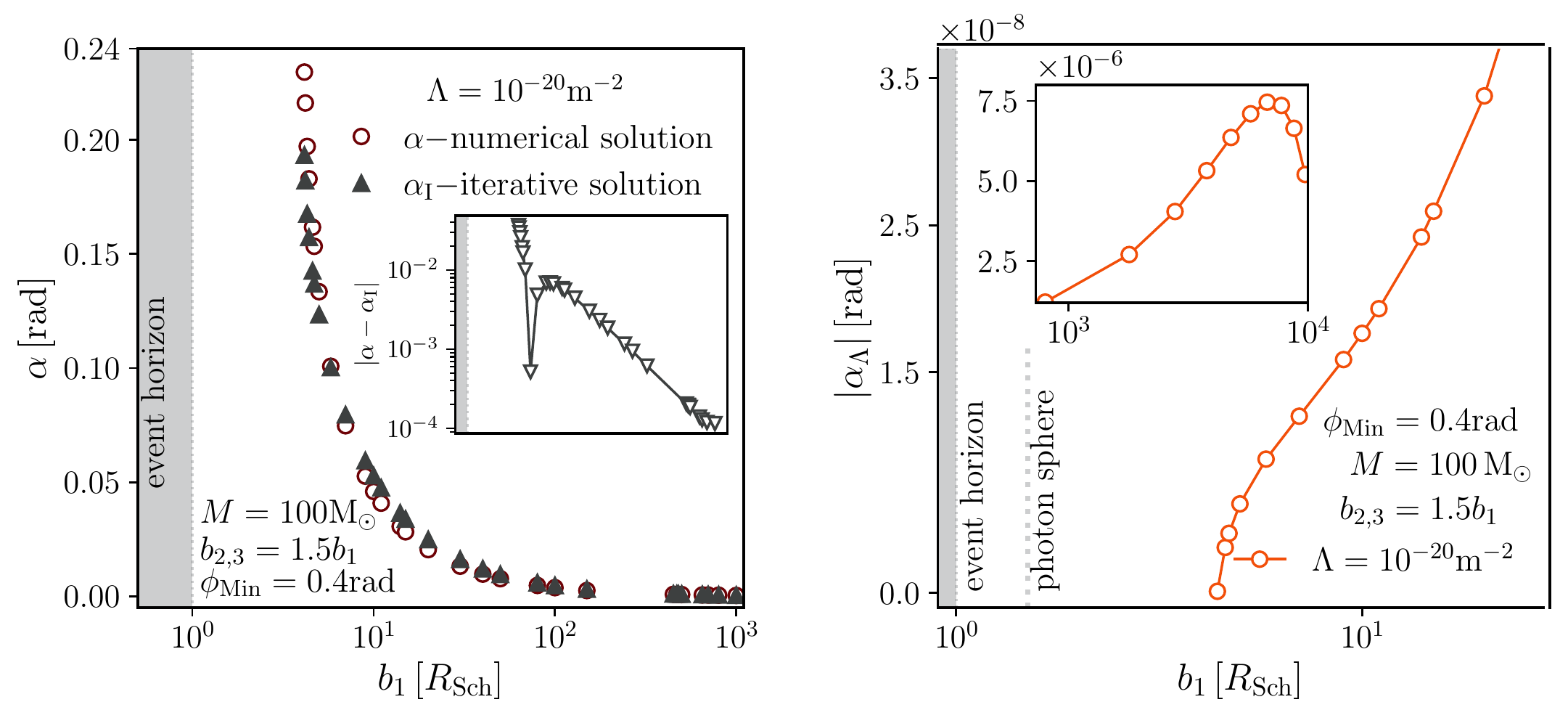}
	\caption{{Influence of the impact parameter $b_1$ in the Horndeski spacetime}. The left panel shows that $\alpha$ vanishes as $b_1$ increases. The inset indicates that the iterative solution~(\ref{Eq:AngHorndeski}) is a poor approximation in the strong field regime. The right panel exhibits how the $\alpha_{\Lambda}$ contribution grows up to a certain value (shown in the inset) and then decreases to zero. Details are explained in the main text.}\label{Fig:H2}
\end{figure*}

The angular difference for configurations located at different distances from the mass source (i.e. for varying $b_1$) are shown in the left panel of figure~\ref{Fig:H2}. From these results, we see that the iterative solution~(\ref{Eq:AngHorndeski}) is a good approximation when the triangular configuration is far from the regime of strong gravity. In particular, for the mass $M=100 \mathrm{M}_\odot$ used in this figure, the inset confirms that the iterative solution is a good approximation for intermediate and large values of $b_{1}$ (as multiples of the Schwarzschild radius). For the regime of strong gravity, the iterative solution~(\ref{Eq:AngHorndeski}) is not a suitable choice. However, one can adopt the approach explored in Ref.~\cite{Chagoya:2020bqz} to address this particular region.

\begin{figure*}[t]
	\centering	
    \includegraphics[width=15.5cm]{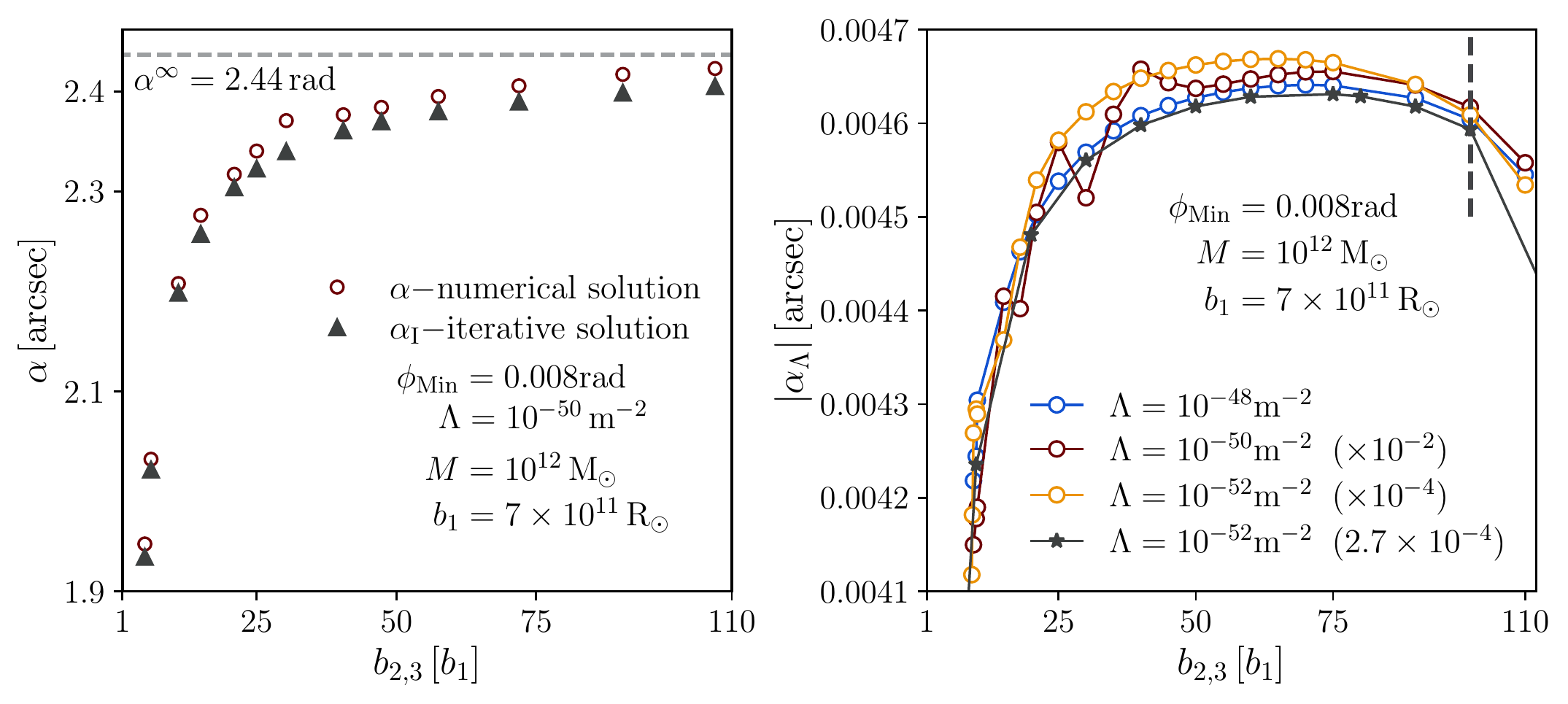}
	\caption{Influence of the impact parameters $b_2, b_3$ (abbreviated as $b_{2,3}$) in the Horndeski spacetime. The left panel exhibits that the Schwarzschild deflection angle $\alpha^{\infty}$ is recovered for a large value of $b_{2,3}$. Similarly to the Kottler spacetime, this happens for small $\phi_{\text{Min}}$. The right panel shows the change of $|\alpha_{\Lambda}|$ with respect to $b_{2,3}$ for different values of $\Lambda$. For comparison, we include the Kottler results (dark star marker). The details are discussed in the main text. The values of $\alpha_\Lambda$ in the plot are to be weighted by the respective multiplicative factor, $10^{-2}$ or $\sim10^{-4}$, shown in the legend.
 \label{Fig:H1}}
\end{figure*}

Now we analyse the effect of varying $b_2=b_3$ on the angular difference. The left panel of fig.~\ref{Fig:H1} shows $\alpha$ for a system with $M=10^{12} \mathrm{M}_\odot$. The deflection angle $\alpha^{\infty}$ is recovered for $b_2=b_3\approx110\times b_1$ with $b_1=7\times10^{11} \mathrm{R}_\odot$. Note that for this calculation the value of $\Lambda$ is larger than the observational value of the cosmological constant.

Concerning the influence of $\Lambda$ on the angular difference $\alpha$, this is encoded in
\begin{align}
    \alpha_{\Lambda}:=\alpha-\alpha_{M}  \approx \alpha_{\Lambda}^{\mathcal{O}(\Lambda)}+\alpha_{\Lambda}^{\mathcal{O}(\Lambda^2)},\label{LamContr}
\end{align}
where $\alpha_{M}$ is the angular difference computed taking $\Lambda=0$, and $\alpha_{\Lambda}^{\mathcal{O}(\Lambda^n)}$
are the contributions of $\Lambda$ to the 
iterative solution~(\ref{Eq:AngHorndeski}) at order $n$ in $\Lambda$. 
Depending on whether we are working with numerical
or iterative solutions, we use the exact expression for $\alpha_\Lambda$ or its approximated form.
The main difference with respect to the Kottler solution is that the contribution from $\Lambda$ is negative. Table~\ref{tabla_2} shows a comparison of $\alpha_\Lambda$ in Kottler and Horndeski spacetimes with a mass of $M=10^{12} \mathrm{M}_\odot$, impact parameters $b_1=7.0\times 10^{11} \mathrm{R}_\odot, b_{2}=b_{3}=100\times b_1$, and $\phi_{\mathrm{Min}}=0.008 \text{rad}$.
For Horndeski, we consider different values of
$\Lambda$. If $\Lambda=10^{-48} \text{m}^{-2}$ the contribution of $\alpha_{\Lambda}$ to $\alpha$ becomes relevant -- around $-0.2\%$ --, which opens the possibility of using the angular difference to constrain theories of gravity. Our results seem to indicate that the angular difference decreases with the increase of $\Lambda$.

Another interesting question is how the impact parameter $b_1$ affects the contribution of $\Lambda$ to the angular difference. The right panel in fig.~\ref{Fig:H2} shows the change in $\alpha_{\Lambda}$ when we move the triangle away from an astrophysical gravity source of $M=100\, \mathrm{M}_\odot$ for $\Lambda=10^{-20} \,\text{m}^{-2}$ and $b_{2, 3}=1.5 b_1$ ($b_{2}, b_{3}$ is abbreviated as $b_{2,3}$). As expected, the contribution grows when the $b_1$ value is increased, which is a consequence of the triangle moving away from the gravitational source reducing its contribution to the angular difference $\alpha$. This growth stops at a sufficiently large value of $b_1$, in this case, $b_1\approx 8250 R_\mathrm{Sch}$; at this point, both triangular configurations -- the background and Horndeski-- are very similar. If the parameter $b_1$ continues to increase, both triangles become almost equal, and $\alpha=0$. Since asymptotically $\alpha_M\to 0$, we conclude that $\alpha_\Lambda\to 0$ as well. This is consistent with the asymptotic behaviour seen in the left panel of figure~\ref{Fig:H2}. On the other hand, as $b_1$ gets smaller, $\alpha_\Lambda$ decreases. This suggests that orbits near the photon sphere are similar to the orbits that one would find in Schwarzschild spacetime. This is explored further in section~\ref{Sec:IV} 
for the Horndeski solution~(\ref{Eq:HorndMetric}).

Now let us see how the impact parameters $b_2, b_3$ affect the contribution of $\Lambda$ to the angular difference. The right panel in figure~\ref{Fig:H1} shows the change in $\abs{\alpha_{\Lambda}}$  when $b_2=b_3$ are increasing for a source of $M=10^{12}\, \mathrm{M}_\odot$ and a trajectory with impact parameter $b_1=7\times10^{11} \mathrm{R}_\odot$. In agreement with the results of  table~\ref{tabla_2}, a change in the order of $\Lambda$ leads to a corresponding change in $\alpha_{\Lambda}$. For instance, when $\Lambda=10^{-52} \;\text{m}^{-2}$, the contribution is $10^{-2}$ times smaller compared to when $\Lambda=10^{-50} \;\text{m}^{-2}$ and $10^{-4}$ times smaller than for $\Lambda=10^{-48} \;\text{m}^{-2}$. For comparison, if $b_2=b_3=100\times b_1$ (vertical dashed line in the right panel of the figure~\ref{Fig:H1}) we have for $\Lambda=10^{-48} \;\text{m}^{-2}$ that its percentage contribution is $27104$ times higher than the corresponding $\Lambda$ contribution in the Kottler spacetime (see table~\ref{tabla_2}). Furthermore, note that $\abs{\alpha_{\Lambda}}$ as a function of $b_{2,3}$ is concave downwards with a maximum in the range $50-75$ times $b_1$. This behaviour is a consequence of: first, for small values of $b_{2}=b_{3}$ the triangles mainly capture information of the gravitational source, i.e., the geodesics $C^{(1)}_{p}$ are mainly determined by the dominant term $2GM/r$ in~(\ref{Eq:HorndMetric}). Second, for intermediate values of $b_{2,3}$ their geodesics $C^{(1)}_{2,3}$ are able to capture the information coming from the $\Lambda$ contribution in~(\ref{Eq:HorndMetric}). Finally, for large values of $b_{2,3}$, we have that $C^{(1)}_{2, 3}$ is less affected by the gravitational source. This means that they are closer to the background geodesics $C^{(2)}_{2, 3}$, which implies that the largest contribution to the angular difference comes from the lower vertices -- dominated by the $2GM/r$ term. This particular behavior opens up the possibility of using the triangular configurations to determine either the mass or cosmological constant contributions to $\alpha$, depending on the position of the upper vertex.

To conclude this section, we emphasise that cosmological studies provide strong constraints on the observed cosmological constant $\Lambda$. In $\Lambda$-CDM -- where the bare and observed cosmological constants coincide -- these studies demand an extremely small value of $\Lambda$. Our results indicate that in order to detect the effect of such a small $\Lambda$ on the angular difference, sources with significant masses
are required.

\renewcommand{\tabcolsep}{10pt}
\renewcommand {\arraystretch}{1.5}
\begin{table*}[t]
	\centering
	\begin{tabular}{|c@{\hskip .08 in}| c@{\hskip .2 in} c@{\hskip .2 in} c@{\hskip .2 in} c@{\hskip .2 in}|}
    \hline
     \multicolumn{5}{|l|}{\textbf{Config.}: $M=10^{12}\mathrm{M}_\odot,\;\; b_1=7.0\times 10^{11} \mathrm{R}_\odot,\;\; b_{2}=b_{3}=100\times b_1,\;\; \phi_{\mathrm{Min}}=0.008 \text{rad}$} \\[0.1cm] \hline 
		Spacetime &  $\Lambda\,[\text{m}^{-2}]$ & $\alpha\,[\text{arcsec}]$ &  $\alpha_{\Lambda}\,[\text{arcsec}]$ & $\%-$contribution \\[0.1cm]
		 \hline
		 Kottler eq.~(\ref{Eq:KottlerMet}) & $1.1\times 10^{-52}$ & $2.4024$& $1.7013\times10^{-7}$ & $7.0815\times10^{-6}$ \\[0.1cm]
		 Horndeski eq.~(\ref{Eq:HorndMetric}) & $1.1\times 10^{-52}$ & $2.4024$ & $-5.0937\times10^{-7}$ & $-2.1202\times10^{-5}$\\[0.1cm]
        Horndeski eq.~(\ref{Eq:HorndMetric}) & $10^{-50}$ & $2.4024$ & $-4.6165\times10^{-5}$ & $-1.9216\times10^{-3}$\\[0.1cm]
        Horndeski eq.~(\ref{Eq:HorndMetric}) &$10^{-48}$ & $2.3978$ & $-4.6024\times10^{-3}$ & $-1.9194\times10^{-1}$ \\[0.1cm]
		\hline
	\end{tabular}
	\caption{Contributions of $\Lambda$ to the angular difference $\alpha$ for a triangular configuration indicated in the first row. The value of $\alpha_{\Lambda}$ is calculated using eq.~(\ref{LamContr}) for different spacetime solutions. In the last column, the respective percentage contribution is given.
	\label{tabla_2}}
\end{table*}

\section{Black hole shadows} \label{Sec:IV}
In this section, we briefly study the black hole shadow in an idealised astrophysical setup where there is no light source between the black hole and a static observer located at a radius $r_{0}$. To calculate the angular radius of the shadow $\alpha_{\text{sh}}$ we use the conventional methodology (see e.g., Refs~\cite{Synge:1966okc, Perlick:2021aok}). An alternative --geometric-- approach is explored in Ref.~\cite{Qiao:2022jlu} using the optical metric~(\ref{Eq:lenOptMe}). 

The angular radius of the shadow corresponding to the only photon sphere at $u_1=2/(3r_s)$, or equivalently $r_1=b/u_1 = 3GM$, existing in the Kottler spacetime~(\ref{Eq:KottlerMet}) was obtained for a static observer in Ref.~\cite{1983BAICz..34..129S, PhysRevD.60.044006}, 
\begin{align}
    \sin^{2}(\alpha^{\text{Kot}}_{\text{sh}})=\frac{27 G^2 M^2}{r^2_0}\left(1-\frac{2GM}{r_0}-\frac{r_0^2\Lambda}{3}\right)\left(1-9G^2M^2\Lambda\right)^{-1},\label{AngKott}
\end{align}
and for an observer comoving with the cosmic expansion in Ref.~\cite{Perlick:2018iye} (eq.~(36)). Alternatively, it has also been deduced in Refs.~\cite{Roy:2020dyy, Tsupko:2019mfo}.

Now, for the Horndeski spacetime~(\ref{Eq:HorndMetric}) we have from the geodesic equation~(\ref{Eq:GeoHord})
\begin{align}
    \left(\frac{du}{d\phi}\right)^2 & := \tilde f = \bigg(1-r_s u-\frac{1}{u^{2}}\bigg)\bigg(\frac{\Lambda}{u^2}-1\bigg),
\end{align}
where $\tilde f$ is
a combination of the total and potential energies, from which the radii of
circular orbits can be obtained. This $\tilde f$ is sometimes called the ``effective potential''. Note that the first parenthesis in $\tilde{f}$ -- in terms of dimensionless variables~(\ref{eq:DimLessVa}) -- coincides with the effective potential of a Schwarzschild spacetime divided by $u^2$. This means that if we assume $u\neq 0$ we will have the known critical points of the Schwarzschild effective potential, $u_1=u_2=\sqrt{3}, u_3=-\sqrt{3}/2$, and a Horndeski critical point $u_4=\sqrt{\Lambda}$. Now, it is important to remind the reader that the critical points are defined as the solutions of $\tilde f=\tilde f '=0$.\footnote{ These conditions are imposed over the radius coordinate, however, this would be equivalent in terms of $u$, see e.g., section $19$ in Ref.~\cite{Chandrasekhar:1985kt}.} The negative root is discarded and for the rest, we have that its critical impact parameters are: $b_1^{\text{cr}}=b_2^{\text{cr}}=3\sqrt{3}GM$, and $b_4^{\text{cr}}=(\Lambda-2GM\Lambda^{3/2})^{-1/2}$. These values separate the capture and the flyby orbits of the incident light rays. From now on we will focus on roots $u_1, u_2$ since, as can be seen from~(\ref{Eq:HorndMetric}), the root $u_4$ corresponds to an event horizon.

Following Ref.~\cite{Perlick:2021aok} one arrives to the equivalent of eq.~(\ref{EqTang}) which for the Horndeski solution~(\ref{Eq:HorndMetric}) in terms of dimensionless variables~(\ref{eq:DimLessVa}) takes the form
\begin{align}
\tan^2\left(\alpha^{\text{Hor}}_{\text{sh}}\right) &=\left(1-\frac{\Lambda}{u^2}\right)\left(1-r_s u\right)\left(\frac{u}{u'}\right)^2\Bigg|_{u_{0}}=\frac{1-r_s u_0}{1-u_0^2+r_s u_0^3}u^2_0,\label{EqAng}
\end{align}
where $u_0=b/r_0$ is the dimensionless observer position. At this point, it is necessary to carefully choose the branch of the solution of the tangent function, see Ref.~\cite{Perlick:2021aok} for a discussion of this. Using the trigonometric property $\sin^2(x)=\tan^2(x)/(1-\tan^2(x))$ we rewrite~(\ref{EqAng}) in a more compact and usual form
\begin{align}
\sin^2\left(\alpha^{\text{Hor}}_{\text{sh}}\right) &=u^2_0-r_s u^3_0=\frac{b^2}{r^2_0}\left(1-\frac{2GM}{r_0}\right),\label{EqAng2}
\end{align}
where in the last step we restored the physical variables (with $\hbar=c=1$). From~(\ref{EqAng2}), we observe that, in contrast to the Kottler result, the presence of $\Lambda$ does not directly influence the expression for $\alpha^{\text{Hor}}_{\text{sh}}$ in the Horndeski case.
Finally, the angular radius of the shadow corresponding to the photon sphere (critical points) $u_1, u_2$ is 
\begin{align}
\sin^2\left(\alpha^{\text{Hor}}_{\text{sh}}\right) &=\frac{27 G^2 M^2}{r^2_0}\left(1-\frac{2GM}{r_0}\right),\label{EqAng3}
\end{align}
which is nothing more than the well-known result for the Schwarzschild spacetime~\cite{Synge:1966okc}.\footnote{ For more complex metrics, the tetrad approach offers an alternative way to determine the angular radius of a shadow. For additional details regarding the tetrad approach, see Refs.~\cite{Grenzebach_2015, Perlick:2021aok}.} For $r_0\gg M$ the angular radius reduced to
\begin{align}
    \alpha^{\text{Hor}}_{\text{sh}}\approx 3\sqrt{3} G M/r_0.
\end{align} 

Finally, using the results reported in Ref.~\cite{Bisnovatyi-Kogan:2018vxl} for a Schwarzschild black hole, we can compute the shadow of the Horndeski solution seen by an observer comoving with the cosmic expansion. For this, we need to fulfill that: i) the cosmological parameters $H_0, \Omega_{i0}$,  where $i$ represents the components of the matter density of the Universe, are close to the $\Lambda-$CDM model,\footnote{ This is necessary to guarantee a viable cosmological evolution. As proven in Refs.~\cite{Bellini:2015xja, Noller:2018wyv}, for Horndeski gravity, the actual cosmological datasets return strong constraints on any possible deviations from $\Lambda-$CDM.}
ii) the observer is at a distance $r_0\gg r_s$, iii) the expansion of spacetime only manifests itself on large --cosmological-- scales. If these conditions are met, then the angular radius of the shadow can be approximated to~\cite{Bisnovatyi-Kogan:2018vxl}
\begin{align}
    \alpha^{\text{Hor}}_{\text{sh}}\approx 3\sqrt{3} G M\frac{H_0 (1+z)}{Int(z)},
\end{align}
where the shadow is at redshift $z$, $H_0$ is the present day value of the Hubble parameter and 
\begin{align}
    Int(z)=\int_{0}^{z}\left(\Omega_{m 0}(1+\tilde{z})^3+\Omega_{r 0}(1+\tilde{z})^4+\Omega_{\Lambda 0}\right)^{-1/2} d\tilde{z},\nonumber
\end{align}
with $\Omega_{m 0}, \Omega_{r 0}, \Omega_{\Lambda 0}$ the present-day values of the density parameters for matter, radiation, and dark energy respectively. For the case $z\ll1 $ we have that $Int(z)\approx z$ and 
\begin{align}
\alpha^{\text{Hor}}_{\text{sh}}\approx 3\sqrt{3} G M\frac{H_0}{z}.
\end{align}

\section{Conclusions}\label{Sec:VI}

We revisited H. Arakida's proposal for computing the deflection angle in non-asymptotically flat spacetimes. This proposal is based on the Gauss-Bonnet theorem and on the difference between the sums of internal angles of two triangles, (\ref{Eq:TotalAngl}), one in the spacetime under consideration, and another one in a background spacetime with zero mass. This allows one to locate the light source or observer at a finite distance. However, the result of eq.~(\ref{Eq:TotalAngl}) does not always coincide with the deflection angle that one would compute considering the trajectory of a single light ray. This is clear from the fact that eq.~(\ref{Eq:TotalAngl}) depends on the impact parameters of three light rays.  For this reason, we reinterpret eq.~(\ref{Eq:TotalAngl}) as an angular difference. The expected result for the deflection angle is recovered in a specific limit.
	
We analysed two non-asymptotically flat spacetimes: a solution to GR, the well-known Kottler spacetime, eq.~(\ref{Eq:KottlerMet}), and a solution to Horndeski gravity, eq.~(\ref{Eq:HorndMetric}). For both spacetimes, we verified -- analytically and numerically -- that eq.~(\ref{Eq:TotalAngl}) effectively removes the pure background contribution to the angular difference. If the top vertex of the triangle is sent much farther away than the bottom geodesic, one recovers the deflection angle for the bottom geodesic subject only to the influence of the mass, while if the whole triangular array is placed far away from the massive source, the angular difference vanishes. These observations support the validity of Arakida's proposal for calculating the deflection angle due to a massive source in non-asymptotically flat spacetimes, using the triangular array described above.
	
Furthermore, we unveiled a feature that has not been emphasised in the literature: for triangular arrays where all vertices are kept at a finite distance, but not too close to the massive source, it is possible to extract the influence of $\Lambda$ on the angular difference from observations performed at different distances from the mass source. Analytically, it can be seen that the contributions of $\Lambda$ appear weighted by a power of $M$, which means that this factor modulates the magnitude of the corrections due to $\Lambda$. In order for the $\Lambda$ contribution to be within the observational sensitivity of future missions, we need very intense gravitational sources. Table~\ref{tabla_1} provides some illustrative numbers.

On the other hand, if the triangular array is too close to the massive source we find that the contribution of $\Lambda$ decreases. Indeed, if the massive source has a photon sphere, at that location the Horndeski solution is effectively stealth in the sense that, due to the structure of the metric, the critical impact parameters, photon sphere radius, and angular radius of the black hole shadow exactly coincide with those of a Schwarzschild black hole.

Despite the similarities between both solutions, there are marked differences. In the Kottler solution, the $\Lambda$ contribution to the angular difference is constructive but extremely small. As discussed above, in order for the $\Lambda$ contribution to be within the observational sensitivity of future missions, we need very intense gravitational sources. On the other hand, for the Horndeski solution, the modification to GR implies that $\Lambda$  is not constrained --necessarily-- to the value of the observed Cosmological constant. Our results indicate that in this case, the Horndeski term contributes negatively to the angular difference, and if it is allowed to be sufficiently large ($\Lambda$ $\sim 10^{-48}-10^{-50} \text{m}^{-2}$) its effect on the angular difference could be within the observational range of future missions (see table~\ref{tabla_2}). 

We would like to note that the fact that the angular difference is equivalent to the deflection angle in a certain limit does not mean that it captures the same information as the RI proposal. Both proposals are a physical consequence of the light bending, but they encode the background contribution differently. The first encodes the massive source and $\Lambda$ contributions over a light ray, while the second encodes the effect of these over three rays that do not  ``feel''  the same effects. This opens up the possibility of using an appropriate choice of the triangular parameters (in our manuscript, these are the impact parameters $b_1, b_2$) to determine the ratio between both contributions in different directions at the same time. The possibility of changing the location and the characteristics of the triangular configurations would allow us to map the gravitational field generated by a massive object, and study for example the isotropy, as well as to tabulate the contribution of $\Lambda$  to the angular difference in different regions around the source.

We end this work with a few comments regarding the physical possibilities of our results. The angular difference is in principle capable of capturing both the effects of the central mass and of the background spacetime, depending on the triangular array used. If we measure the angular difference of a triangular configuration closer to a strong gravitational source, and then repeat the same measurement moving away, where the background contribution is dominant, we could compare the results with the prediction given for different spacetimes, such as the Kottler or Horndeski solution, and determine, for example, the value of $\Lambda$ or the concordant gravitational model. Additionally, by using the predicted angular difference corresponding to the visible matter of a gravitational source, we could constrain the dark matter contribution (or effects of modifications to GR) by comparing the measurements with the theoretical prediction. 
The construction of a triangular configuration for these scenarios is beyond our scope. 
On the other hand, the use of the Solar System as a laboratory is not completely ruled out. Triangular configurations similar to the ones studied in this work could be possible in future missions e.g., LATOR~\cite{Turyshev:2003wt}, ASTROD-GW~\cite{Ni:2012eh}, LISA~\cite{Bayle:2022hvs, Amaro_Seoane_2023} (for other mission proposals see table 1 in Ref.~\cite{Ni:2016wcv}), where the laser-beam baselines can be aboard these missions. However, the possibility of using these as a detector of the $\Lambda$ contribution  would be beyond our current technical possibilities. For example, for LISA, the angular difference will be around a micro-arcsec, with a $\Lambda$ contribution of $10^{-30}$ arcsec, while for a triangular configuration with impact parameters $b_{2}=b_{3}=1$ AU and $b_{0}=1.5$ AU, one gets $\alpha\sim 10^{-3}$ arcsec, but a $\Lambda$ contribution of order $10^{-29}$ arcsec. In both cases, the expected $\Lambda$ contributions are not within our observational possibilities. Nevertheless, in this context, the angular deflection can be used to study effects coming from other solutions where the contributions do not depend only on the product of $M\Lambda$ or to study properties of the gravitational source. This idea will be explored in future works, as well as the influence of the structural shape of the triangular configuration on the angular difference.
Another direction for extending our results is to consider geodesics of massive particles, either by applying the numerical techniques developed in this paper or by introducing an effective optical metric as in Ref.~\cite{PhysRevD.97.124016} and using the Gauss-Bonnet theorem.

\subsection*{Acknowledgments}
A.A.R. acknowledges funding from a postdoctoral fellowship from ``Estancias Posdoctorales por México para la Formación y Consolidación de las y los Investigadores por México''.
F.S. and J.C. are supported by CONACyT/DCF/320821, and B.R. by CONACyT/932448.

\appendix
\addtocontents{toc}{\fixappendix}
\section{Technical results}\label{Sec:Appendix}
This Appendix presents technical results corresponding to some mathematical expressions used in the main part of the paper.  

\subsection{Technical results for the Kottler solution}

In section~\ref{Sec:II.3} we present the result of the surface integral eq.~(\ref{Eq:IntSurf}),
\begin{align}
    \mathcal{A}^{(C_p, I)}= \Gamma_{0}^{(C_p)}+\frac{2GM}{b_p} \Gamma_{1}^{(C_p)}+\left(\frac{2GM}{b_p}\right)^2 \Gamma_{2}^{(C_p)}+\mathcal{O}\left(\frac{2GM}{b_p}\right)^{3},
\end{align}
where the $\Gamma_{i}^{(C_p)}$ are:
{\small  
  \setlength{\abovedisplayskip}{6pt}
  \setlength{\belowdisplayskip}{\abovedisplayskip}
  \setlength{\abovedisplayshortskip}{0pt}
  \setlength{\belowdisplayshortskip}{3pt}
\begin{subequations}\label{GammaExp}
    \begin{align}
        \Gamma_{0}^{(C_p)}&:=-\phi+\frac{\cot{(\phi-\delta_p)}}{6}\Lambda b_{p}^{2}+\frac{\csc^{4}{(\phi-\delta_p)}\sin{(4\phi-4\delta_p)}}{288}\Lambda^2 b_{p}^{4}+\mathcal{O}\left(\Lambda b_{p}^{2}\right)^3,\\
        \Gamma_{1}^{(C_p)}&:=-\cos{(\phi-\delta_p)}-\frac{\cot{(\phi-\delta_p)}\csc{(\phi-\delta_p)}}{12}\bigg(3\left[1-\frac{\cos{(2\phi-2\delta_p)}}{3}\right]\Lambda b_{p}^{2}\nonumber\\ 
        &+\frac{7 \cot^2{(\phi-\delta_p)}}{12}\bigg[1-\frac{\cos{(2\phi-2\delta_p)}}{7}\bigg]\Lambda^2 b_{p}^{4}\bigg)+\mathcal{O}\left(\Lambda b_{p}^{2}\right)^3,\\
        \Gamma_{2}^{(C_p)}&:=\frac{30 (\phi-\delta_p)}{32}\bigg[1+\frac{\sin{(2\phi-2\delta_p)}}{30(\phi-\delta_p)}\bigg]-\frac{1}{192}\bigg[ 30\bigg(1-\frac{32 \csc^{2}{(\phi-\delta_p)}}{30}\bigg)\cot{(\phi-\delta_p)}\nonumber\\
        &-60(\phi-\delta_p)\bigg(1+\frac{\sin{(2\phi-2\delta_p)}}{30(\phi-\delta_p)}\bigg)+15\Phi\csc^2{(\phi-\delta_p)}\bigg]\Lambda b_{p}^{2}-\frac{30\csc^{4}(\phi-\delta_p)\Phi}{1536}\bigg[1\nonumber\\
        &-\frac{97\cot{(\phi-\delta_p)}}{30\Phi}-\frac{31\cos{(3\phi-3\delta_p)}\csc{(\phi-\delta_p)}}{30\Phi}\bigg] \Lambda^2 b_{p}^{4}+\mathcal{O}\left(\Lambda b_{p}^{2}\right)^3.
    \end{align}
\end{subequations}
}

From this expression, and using an iterative approach, we calculated the angular difference $\alpha$ for the Kottler solution~(\ref{Eq:KottlerMet}) whose result is reported in equation~(\ref{Eq:AngSchDSitter}) and reads as follows:
{\small  
  \setlength{\abovedisplayskip}{6pt}
  \setlength{\belowdisplayskip}{\abovedisplayskip}
  \setlength{\abovedisplayshortskip}{0pt}
  \setlength{\belowdisplayshortskip}{3pt}
\begin{align}
    \alpha&= \Bigg|4GM\bigg\lbrace \frac{\cos([\delta+\phi_2])+\sin(\delta)}{b_2}-\frac{\cos(\phi_2)}{b_1}+\frac{\Lambda}{24}\alpha_{11}+\frac{\Lambda^2}{144}\alpha_{12}\bigg\rbrace+\frac{G^2M^2}{4}\bigg\lbrace\frac{1}{b_1^2}\bigg[\sin(2\phi_2)\nonumber\\
    &-15\pi\bigg(1-\frac{b_1^2}{b_2^2}\left[1-\frac{2\phi_2}{\pi}\right]-\frac{2\phi_2}{\pi}\bigg)-\frac{2 b_1^2}{b_2^2}\cos(\phi_2)\sin(\phi_2+2\delta)\bigg]+\frac{\Lambda}{6}\alpha_{21}+\frac{\Lambda^2}{96}\alpha_{22}\bigg\rbrace\Bigg|.
\end{align}
}
where the terms $\alpha_{ij}$ are
{\small  
  \setlength{\abovedisplayskip}{6pt}
  \setlength{\belowdisplayskip}{\abovedisplayskip}
  \setlength{\abovedisplayshortskip}{0pt}
  \setlength{\belowdisplayshortskip}{3pt}
\begin{subequations}\label{AlphaKot}
\begin{align}
\alpha_{11}&=-4b_1\big[\cos{(\phi_2)}+\cot{(\phi_2)}\csc{(\phi_2)}\big]+b_2\bigg[\csc^2{\bigg(\frac{\phi_2+\delta}{2}\bigg)}-\frac{2}{1+\cos{(\phi_2+\delta)}}\nonumber\\
    &+4\sin{(\delta)[1-\sin{(\phi_2)}-4\sec{(\delta)\tan{(\delta)}}]+4\cos{(\delta)}\cos{(\phi_2)}}\bigg],\\
    \alpha_{12}&=b^{3}_{1}(\cos{(2\phi_2)}-7)\cot^3{(\phi_2)}\csc{(\phi_2)}+b^{3}_{2}\bigg[(7+\cos{(2\delta)})\sec{(\delta)\tan^3{(\delta)}}\nonumber\\
    &+(7-\cos{[2(\phi_2+\delta)]})\cot^3{(\phi_2+\delta)}\csc{(\phi_2+\delta)}\bigg],\\
    \alpha_{21}&=30\cot{(\phi_2+\delta)}\bigg(1-\frac{32\csc^2{(\phi_2+\delta)}}{30}\bigg)-30\cot{(\phi_2)}\bigg(1-\frac{32\csc^2{(\phi_2)}}{30}\bigg)\nonumber\\
    &+30\delta\sec^2{(\delta)}[1+(\pi-2(\phi_2+\delta))\csc^2{(\phi_2+\delta)}-(\pi-2\phi_2)\csc^2{(\phi_2)}]\nonumber\\
    &+28\bigg(1-\frac{\cos{(2\delta)}}{14}-\frac{\cos{(2\phi_2)}}{14}-\frac{\cos{[2(\phi_2+\delta)]}}{14}-\frac{8\sec^{2}{(\delta)}}{7}\bigg)\tan{(\delta)},\\
    \alpha_{22}&=2b^{2}_{1}(60\phi_2+97\cot{(\phi_2)}+31\cos{(3\phi_2)}\csc{(\phi_2)}-30\pi)\csc^4{(\phi_2)}\nonumber\\
    &+12b^{2}_{2}\bigg[(5[\pi-2(\delta+\phi_2)]-11\cot{(\phi_2+\delta)})\csc^{4}{(\phi_2+\delta)})\nonumber\\
    &-\frac{31\csc^{6}{(\phi_2+\delta)}\sin{[4(\phi_2+\delta)]}}{12}+\frac{\sec^{5}{(\delta)}}{6}(60\delta\cos{(\delta)}-97\sin{(\delta)}\nonumber\\
    &+31\sin{(3\delta)})\bigg].
\end{align}
\end{subequations}
}
\subsection{Technical results for the Horndeski solution}

In section~\ref{Sec:II.4} we presented eq.~(\ref{Eq:uSerieHord}) that corresponds to the first term of the full iterative solution:
{\small  
  \setlength{\abovedisplayskip}{6pt}
  \setlength{\belowdisplayskip}{\abovedisplayskip}
  \setlength{\abovedisplayshortskip}{0pt}
  \setlength{\belowdisplayshortskip}{3pt}
\begin{align}
u^{(C_p, I)} &= \sin(\phi-\delta_p)\left(1+\frac{\cot^2{(\phi-\delta_p)}}{2}\bar{\Lambda}-\frac{\cot^4{(\phi-\delta_p)}}{8}\bar{\Lambda}^2\right)+\frac{3}{4}\bigg\lbrace 1+\frac{\cos{(2\phi-2\delta_p)}}{3}\nonumber\\
&-\cot^2{(\phi-\delta_p)}\left[1-\frac{\cos{(2\phi-2\delta_p)}}{3}\right]\bar{\Lambda}+\frac{7 \cot^2{(\phi-\delta_p)}}{16 \sin^2{(\phi-\delta_p)}}\bigg[1+\frac{4\sin^4{(\phi-\delta_p)}}{7\cos{(\phi-\delta_p)}}\nonumber\\
&\ln{\left(\tan{\left[\frac{\phi}{2}-\frac{\delta_p}{2}\right]}\right)}+\frac{\cos{(3\phi-3\delta_p)}}{7\cos{(\phi-\delta_p)}}\bigg]\bar{\Lambda}^2\bigg\rbrace r_s+\frac{37}{64}\bigg\lbrace \sin{(\phi-\delta_p)}\bigg[1-\frac{3\sin{(3\phi-3\delta_p)}}{37\sin{(\phi-\delta_p)}}\nonumber\\
&+\frac{30\Phi}{37}\bigg]+\frac{123}{148}\frac{\cot^{2}{(\phi-\delta_p)}}{\sin{(\phi-\delta_p)}}\Bigg[1-\frac{29\Phi\tan{(\phi-\delta_p)}}{41}\bigg(1-\frac{3\sin{(3\phi-3\delta_p)}}{29\sin{(\phi-\delta_p)}}\bigg)\nonumber\\
&+\sum_{j=1}^{2}\frac{\cos{([2j+1][\phi-\delta_p])}}{\cos{(\phi-\delta_p)}}\mathrm{V}_{j}^{(1)}\bigg]\bar{\Lambda}-\frac{4669}{4736}\frac{\cot^{2}{(\phi-\delta_p)}}{\sin^{3}{(\phi-\delta_p)}}\bigg[1-\frac{1536}{4669}\frac{\sin^{6}{(\phi-\delta_p)}}{\cos{(\phi-\delta_p)}}\nonumber\\
&\ln{\bigg(\cot{\bigg[\frac{\phi}{2}-\frac{\delta_p}{2}\bigg]}\bigg)}-\frac{444 \Phi}{4669}\tan{(\phi-\delta_p)}\bigg(1+\sum_{j=1}^{2}(-1)^{j-1}\cos{(2i[\phi-\delta_p])}\mathrm{V}_{j}^{(2)}\bigg)\nonumber\\
&+\sum_{j=1}^{3}\frac{\cos{([2j+1][\phi-\delta_p])}}{\cos{(\phi-\delta_p)}}\mathrm{V}_{j}^{(3)}\bigg]\bar{\Lambda}^2\bigg\rbrace r_s^2+\mathcal{O}(r_s^{3}, \bar{\Lambda}^3),\label{AEq:uSerieHord}
\end{align}
}
where $\Phi:=\pi-2(\phi-\delta_p)$, and we defined the arrays
\begin{align}
   [V_j^{(1)}]:=&\left(\frac{1}{246}, \frac{3}{82}\right),\quad [V_j^{(2)}]:=\left(\frac{92}{37}, \frac{9}{37}\right),\quad [V_j^{(3)}]:=\left(\frac{217}{667}, -\frac{5}{667}, -\frac{9}{4669}\right). \nonumber 
\end{align}
Using this result, we computed the surface integral eq.~(\ref{Eq:IntSurf}), giving the terms $\mathcal{A}^{(C_p, I)}$
\begin{align}
    \mathcal{A}^{(C_p, I)}= \Gamma_{0}^{(C_p)}+\frac{2GM}{b_p} \Gamma_{1}^{(C_p)}+\left(\frac{2GM}{b_p}\right)^2 \Gamma_{2}^{(C_p)}+\mathcal{O}\left(\frac{2GM}{b_p}\right)^{3},
\end{align}
with
{\small  
  \setlength{\abovedisplayskip}{6pt}
  \setlength{\belowdisplayskip}{\abovedisplayskip}
  \setlength{\abovedisplayshortskip}{0pt}
  \setlength{\belowdisplayshortskip}{3pt}
\begin{subequations}\label{HorCp}
    \begin{align}
        \Gamma_{0}^{(C_p)}&:=-\phi-\frac{\cot{(\phi-\delta_p)}}{2}\Lambda b_{p}^{2}+\frac{\csc^{4}{(\phi-\delta_p)}\sin{(4[\phi-\delta_p])}}{32}\Lambda^2 b_{p}^{4}+\mathcal{O}\left(\Lambda b_{p}^{2}\right)^3,\\
        \Gamma_{1}^{(C_p)}&:=-\cos{(\phi-\delta_p)}+\frac{1}{8}\bigg[4\cos{(\phi-\delta_p)}-\frac{2}{1+\cos{(\phi-\delta_p)}}+\csc^2{\left(\frac{\phi}{2}-\frac{\delta_p}{2}\right)}\bigg]\Lambda b_{p}^{2}\nonumber\\ &+\frac{1}{128}\bigg[\ln{\bigg(\cot^{24}{\bigg[\frac{\phi}{2}-\frac{\delta_p}{2}\bigg]}\bigg)}+4\csc^4{\left(\frac{\phi}{2}-\frac{\delta_p}{2}\right)}\bigg(1-\frac{7\cos{(\phi-\delta_p)}}{4}\bigg)\nonumber\\
        &-\frac{28}{1+\cos{(\phi-\delta_p)}}+3\sec^{4}{\left(\frac{\phi}{2}-\frac{\delta_p}{2}\right)}+16\cos{(\phi-\delta_p)}\bigg]\Lambda^{2} b_{p}^{4}+\mathcal{O}\left(\Lambda b_{p}^{2}\right)^3,
    \end{align}
    \begin{align}
        \Gamma_{2}^{(C_p)}&:=\frac{\sin{(2[\phi-\delta_p])}}{32}\bigg[1+\frac{30(\phi-\delta_p)}{\sin{(2[\phi-\delta_p])}}\bigg]+\frac{1}{64}\bigg[15\Phi\csc^{2}{(\phi-\delta_p)}\nonumber\\
        &+30\cot{(\phi-\delta_p)}\bigg(1-\frac{32 \csc^2{(\phi-\delta_p)}}{30}\bigg)-12\delta_p\bigg(1-\frac{\phi}{\delta_p}+\frac{\sin{(2[\phi-\delta_p])}}{6\delta_p}\bigg)\bigg]\Lambda b_{p}^{2}\nonumber\\
        &+\frac{255 \csc^{5}{(\phi-\delta_p)}}{512}\bigg[\frac{41 \cos{(3[\phi-\delta_p])}}{85}+\frac{32\sin^{6}{(\phi-\delta_p)}}{85}\ln\bigg(\tan{\bigg[\frac{\phi}{2}-\frac{\delta_p}{2}\bigg]}\bigg)\nonumber\\
        &+\frac{2\cos{(5[\phi-\delta_p])}}{85}-\frac{6\Phi\sin{(\phi-\delta_p)}(1+4\cos{(2[\phi-\delta_p])})}{85}+\cos{(\phi-\delta_p)}\bigg]\Lambda^{2} b_{p}^{4}\nonumber\\
        &+\mathcal{O}\left(\Lambda b_{p}^{2}\right)^3.
    \end{align}
\end{subequations}
}
Finally from the above results the angular difference $\alpha$ for the Horndeski gravity~(\ref{Eq:HorndMetric}) solution, is computed and reported in equation~(\ref{Eq:AngHorndeski}), and reads as follows,
\begin{align}
    \alpha&= \Bigg|4GM\bigg\lbrace \frac{\cos([\delta+\phi_2])+\sin(\delta)}{b_2}-\frac{\cos(\phi_2)}{b_1}+\frac{\Lambda}{8}\alpha^{H}_{11}+\frac{\Lambda^2}{128}\alpha^{H}_{12}\bigg\rbrace\nonumber\\
    &+\frac{G^2M^2}{4}\bigg\lbrace\frac{1}{b_1^2}\left[\sin(2\phi_2)-15\pi\bigg(1-\frac{b_1^2}{b_3^2}\left[1-\frac{2\phi_2}{\pi}\right]-\frac{2\phi_2}{\pi}\bigg)-\frac{2 b_1^2}{b_2^2}\cos(\phi_2)\sin(\phi_2+2\delta)\right]\nonumber\\
    &+\frac{\Lambda}{2}\alpha^{H}_{21}+\frac{3\Lambda^2}{4}\alpha^{H}_{22}\bigg\rbrace\Bigg|.
\end{align}
where the $\alpha^{H}_{ij}$ terms are
\begin{subequations}\label{AlphaHornd}
\begin{align}
\alpha^{H}_{11}&=4b_1(1+\csc^2{(\phi_2)})\cos{(\phi_2)}+b_2\bigg[\frac{2}{\cos{(\phi_2+\delta)}-1}+\sec^{2}{\bigg(\frac{\phi_2+\delta}{2}\bigg)}\nonumber\\
&-4\cos{(\delta)}\cos{(\phi_2)}+4\sin{(\delta)}\sin{(\phi_2)}-2(3+\cos{(2\delta)})\sec{(\delta)}\tan{(\delta)}\bigg],\\
\alpha^{H}_{12}&=8 b^{3}_1\bigg[2\cos{(\phi_2)}+7\cot{(\phi_2)}\csc{(\phi_2)}\bigg(1-\frac{6\csc^{2}{(\phi_2)}}{7}\bigg)+\ln{\bigg(\cot^{3}{\bigg[\frac{\phi_2}{2}\bigg]}\bigg)}\bigg]\nonumber\\
&-\frac{b^{3}_{3}}{\sin^{2}{(\phi_2)}\sin^{2}{(\phi_2+\delta)}\cos^{2}{(\delta)}}\bigg[24\ln{\bigg(\cot{\bigg[\frac{\pi+2\delta}{4}\bigg]}\tan{\bigg[\frac{\phi_2}{2}+\frac{\delta}{2}\bigg]}\bigg)}\nonumber\\
&+\csc^2{(\phi_2+\delta)}\sin^{2}{(\phi_2)}\sec^{2}{(\delta)}\bigg\lbrace \sin^{4}{(\phi_2+\delta)}\sin{(5\delta)}\bigg[1+\frac{17\sin{(3\delta)}}{\sin{(5\delta)}}-\frac{32\sin{(\delta)}}{\sin{(5\delta)}}\bigg]\nonumber\\
&-32\cos{(\phi_2)}\cos^5{(\delta)}\bigg[1-\frac{17\cos{(3\phi_2)}}{32\cos{(\phi_2)}}\bigg(1-\frac{34\cos{(2\delta)}}{17}\bigg)-\frac{\cos{(5\phi_2)}}{32\cos{(\phi_2)}}(1-2\cos{(2\delta)}\nonumber\\
&+2\cos{(4\delta)})\bigg]+32\sin{(\phi_2)}\sin{(\delta)}\cos^4{(\delta)}\bigg[1+\frac{17\sin{(3\phi_2)}\sin{(3\delta)}}{32\sin{(\phi_2)}\sin{(\delta)}}\nonumber\\
&-\frac{\sin{(5\phi_2)\sin{(5\delta)}}}{32\sin{(\phi_2)}\sin{(\delta)}}\bigg]\bigg\rbrace \bigg],
\end{align}
\begin{align}
\alpha^{H}_{21}&=-\alpha_{21},\\
\alpha^{H}_{22}&=\frac{b^{2}_{1}\csc^{5}{(\phi_2)}}{2}\bigg[85\cos{(\phi_2)}+41\cos{(3\phi_2)}+2\cos{(5\phi_2)}-32\ln{\bigg(\cot\bigg[\frac{\phi_2}{2}\bigg]\bigg)}\sin^{6}(\phi_2)\nonumber\\
    &+6(\pi-2\phi_2)(\sin{(\phi_2)}-2\sin{(3\phi_2)})\bigg]+b^{2}_{3}\Omega\bigg[1\nonumber\\
    &-\frac{16\cot{(\phi_2+\delta)}}{\Omega}\bigg(1+4\csc^{4}{(\phi_2+\delta)}-\frac{47\csc^{2}{(\phi_2+\delta)}}{8}\bigg)-\frac{6\delta\cos{(\delta)}\sec^{5}{(\delta)}}{\Omega}\bigg(1\nonumber\\
    &+\frac{2\cos{(3\delta)}}{\cos{(\delta)}}+\frac{85\tan{(\delta)}}{12\delta}-\frac{41\sin{(3\delta)}}{12\delta\cos{(\delta)}}+\frac{\sin{(5\delta)}}{6\delta\cos{(\delta)}}\bigg)\nonumber\\
    &+\frac{3(\pi-2(\phi_2+\delta))(1+4\cos{[2(\phi_2+\delta)]})\csc^{4}{(\phi_2+\delta)}}{\Omega}
    \bigg],
\end{align}
\end{subequations}
with 
\begin{align}
\Omega:=-16\bigg(\sin{(\phi_2+\delta)}\ln{\bigg[\tan{\bigg(\frac{\phi_2}{2}+\frac{\delta}{2}\bigg)}\bigg]}+\cos{(\delta)}\ln{\bigg[\cot{\bigg(\frac{\pi+2\delta}{4}\bigg)}\bigg]}\bigg).\nonumber
\end{align}

\section*{References}

\bibliographystyle{unsrt}
\bibliography{refv3}

\begin{thebibliography}{10}

\bibitem{Zubairy2016}
M.~S. Zubairy.
\newblock {A Very Brief History of Light}.
\newblock In M.~D. Al-Amri, M.~El-Gomati, and M.~S. Zubairy, editors, {\em
  {Optics in Our Time}}, pages 3--24. Springer International Publishing, Cham,
  2016.

\bibitem{Eddington1920}
F.~W. Dyson, A.~S. Eddington, and C.~Davidson.
\newblock {IX. A determination of the deflection of light by the sun's
  gravitational field, from observations made at the total eclipse of May 29,
  1919}.
\newblock {\em Philosophical Transactions of the Royal Society of London.
  Series A, Containing Papers of a Mathematical or Physical Character},
  220(571-581):291--333, 1920.

\bibitem{Earman1980}
J.~Earman and C.~Glymour.
\newblock {Relativity and Eclipses: The British Eclipse Expeditions of 1919 and
  Their Predecessors}.
\newblock {\em Historical Studies in the Physical Sciences}, 11(1):49--85, 01
  1980.

\bibitem{Ellis2010GravitationalLA}
R.~S. Ellis.
\newblock {Gravitational lensing: a unique probe of dark matter and dark
  energy}.
\newblock {\em Philos. Trans. A Math. Phys. Eng. Sci.}, 368:967--987, 2010.

\bibitem{Shapiro:2004zz}
S.~S. Shapiro, J.~L. Davis, D.~E. Lebach, and J.~S. Gregory.
\newblock {Measurement of the Solar Gravitational Deflection of Radio Waves
  using Geodetic Very-Long-Baseline Interferometry Data, 1979-1999}.
\newblock {\em Phys. Rev. Lett.}, 92:121101, 2004.

\bibitem{Schmidt_2008}
F.~Schmidt.
\newblock {Weak Lensing Probes of Modified Gravity}.
\newblock {\em Phys. Rev. D}, 78:043002, 2008.

\bibitem{Uzan:2010ri}
J.~P. Uzan.
\newblock {Tests of General Relativity on Astrophysical Scales}.
\newblock {\em Gen. Rel. Grav.}, 42:2219--2246, 2010.

\bibitem{Pratten_2016}
G.~Pratten, D.~Munshi, P.~Valageas, and P.~Brax.
\newblock {3D Weak Lensing: Modified Theories of Gravity}.
\newblock {\em Phys. Rev. D}, 93(10):103524, 2016.

\bibitem{Baker:2019gxo}
T.~Baker et~al.
\newblock {Novel Probes Project: Tests of gravity on astrophysical scales}.
\newblock {\em Rev. Mod. Phys.}, 93(1):015003, 2021.

\bibitem{EventHorizonTelescope:2019ggy}
K.~Akiyama et~al.
\newblock {First M87 Event Horizon Telescope Results. VI. The Shadow and Mass
  of the Central Black Hole}.
\newblock {\em Astrophys. J. Lett.}, 875(1):L6, 2019.

\bibitem{SagittariusA}
Event Horizon~Telescope Collaboration and et~al.
\newblock {First Sagittarius A* Event Horizon Telescope Results. I. The Shadow
  of the Supermassive Black Hole in the Center of the Milky Way}.
\newblock {\em The Astrophysical Journal Letters}, 930(L12), 2022.

\bibitem{Cunha:2017wao}
P.~V.~P. Cunha, J.~A. Font, C.~Herdeiro, E.~Radu, N.~Sanchis-Gual, and
  M.~Zilh\~ao.
\newblock {Lensing and dynamics of ultracompact bosonic stars}.
\newblock {\em Phys. Rev. D}, 96(10):104040, 2017.

\bibitem{Liebling:2012fv}
S.~L. Liebling and C.~Palenzuela.
\newblock {Dynamical boson stars}.
\newblock {\em Living Rev. Rel.}, 26(1):1, 2023.

\bibitem{Alcubierre:2018ahf}
M.~Alcubierre, J.~Barranco, A.~Bernal, J.~C. Degollado, A.~Diez-Tejedor,
  M.~Megevand, D.~N\'u\~nez, and O.~Sarbach.
\newblock {$\ell$-Boson stars}.
\newblock {\em Class. Quant. Grav.}, 35(19):19LT01, 2018.

\bibitem{Alcubierre:2021psa}
M.~Alcubierre, J.~Barranco, A.~Bernal, J.~C. Degollado, A.~Diez-Tejedor,
  V.~Jaramillo, M.~Megevand, D.~N\'u\~nez, and O.~Sarbach.
\newblock {Extreme \ensuremath{\ell}-boson stars}.
\newblock {\em Class. Quant. Grav.}, 39(9):094001, 2022.

\bibitem{Alcubierre:2022rgp}
M.~Alcubierre, J.~Barranco, A.~Bernal, J.~C. Degollado, A.~Diez-Tejedor,
  M.~Megevand, D.~N\'u\~nez, and O.~Sarbach.
\newblock {Boson stars and their relatives in semiclassical gravity}.
\newblock {\em Phys. Rev. D}, 107(4):045017, 2023.

\bibitem{Roque:2023sjl}
A.~A. Roque, E.~C. Nambo, and O.~Sarbach.
\newblock {Radial linear stability of nonrelativistic \ensuremath{\ell}-boson
  stars}.
\newblock {\em Phys. Rev. D}, 107(8):084001, 2023.

\bibitem{Barranco:2021auj}
J.~Barranco, J.~Chagoya, A.~Diez-Tejedor, G.~Niz, and A.~A. Roque.
\newblock {Horndeski stars}.
\newblock {\em JCAP}, 10:022, 2021.

\bibitem{Roque:2021lvr}
A.~A. Roque and L.~A. Ureña-L\'opez.
\newblock {Horndeski fermion\textendash{}boson stars}.
\newblock {\em Class. Quant. Grav.}, 39(4):044001, 2022.

\bibitem{Cardoso:2019rvt}
V.~Cardoso and P.~Pani.
\newblock {Testing the nature of dark compact objects: a status report}.
\newblock {\em Living Rev. Rel.}, 22(1):4, 07 2019.

\bibitem{Misner:1973prb}
C.~W. Misner, K.~S. Thorne, and J.~A. Wheeler.
\newblock {\em {Gravitation}}.
\newblock W. H. Freeman, San Francisco, 1973.

\bibitem{Weinberg:1972kfs}
S.~Weinberg.
\newblock {\em {Gravitation and Cosmology}: {Principles and Applications of the
  General Theory of Relativity}}.
\newblock John Wiley and Sons, New York, 1972.

\bibitem{Perlick:2010zh}
V.~Perlick.
\newblock {Gravitational Lensing from a Spacetime Perspective}.
\newblock {\em Living Rev. Rel.}, 7(9):1433--8351, 12 2004.
\newblock Updated version: arXiv:1010.3416.

\bibitem{doi:10.1098/rspa.1959.0015}
C.~Darwin.
\newblock {The Gravity Field of a Particle}.
\newblock {\em Proceedings of the Royal Society of London Series A},
  249(1257):180--194, January 1959.

\bibitem{1972ApJ...173L.137C}
C.~T. Cunningham and J.~M. Bardeen.
\newblock {The Optical Appearance of a Star Orbiting an Extreme Kerr Black
  Hole}.
\newblock {\em Astrophys. J.}, 173:L137, May 1972.

\bibitem{1979A&A....75..228L}
J.~P. Luminet.
\newblock {Image of a spherical black hole with thin accretion disk}.
\newblock {\em Astron. Astrophys.}, 75:228--235, 1979.

\bibitem{Bozza:2005tg}
V.~Bozza, F.~De~Luca, G.~Scarpetta, and M.~Sereno.
\newblock {Analytic Kerr black hole lensing for equatorial observers in the
  strong deflection limit}.
\newblock {\em Phys. Rev. D}, 72:083003, 2005.

\bibitem{Aazami:2011tu}
A.~B. Aazami, C.~R. Keeton, and A.~O. Petters.
\newblock {Lensing by Kerr Black Holes. I: General Lens Equation and
  Magnification Formula}.
\newblock {\em J. Math. Phys.}, 52:092502, 2011.

\bibitem{Ghosh:2022mka}
S.~Ghosh and A.~Bhattacharyya.
\newblock {Analytical study of gravitational lensing in Kerr-Newman
  black-bounce spacetime}.
\newblock {\em JCAP}, 11:006, 2022.

\bibitem{Bozza_2001}
V.~Bozza, S.~Capozziello, G.~Iovane, and G.~Scarpetta.
\newblock {Strong field limit of black hole gravitational lensing}.
\newblock {\em Gen. Rel. Grav.}, 33:1535--1548, 2001.

\bibitem{Bozza_2002}
V.~Bozza.
\newblock {Gravitational lensing in the strong field limit}.
\newblock {\em Phys. Rev. D}, 66:103001, 2002.

\bibitem{Virbhadra:1998dy}
K.~S. Virbhadra, D.~Narasimha, and S.~M. Chitre.
\newblock {Role of the scalar field in gravitational lensing}.
\newblock {\em Astron. Astrophys.}, 337:1--8, 1998.

\bibitem{Izmailov:2019uhy}
R.~N. Izmailov, R.~K. Karimov, E.~R. Zhdanov, and K.~K. Nandi.
\newblock {Modified gravity black hole lensing observables in weak and strong
  field of gravity}.
\newblock {\em Mon. Not. Roy. Astron. Soc.}, 483(3):3754--3761, 2019.

\bibitem{Chagoya:2020bqz}
J.~Chagoya, C.~Ortiz, B.~Rodr\'\i{}guez, and A.~A. Roque.
\newblock {Strong gravitational lensing by DHOST black holes}.
\newblock {\em Class. Quant. Grav.}, 38(7):075026, 2021.

\bibitem{Kottler1918}
F.~Kottler.
\newblock {{\"U}ber die physikalischen Grundlagen der Einsteinschen
  Gravitationstheorie}.
\newblock {\em Annalen der Physik}, 361(14):401--462, 1918.

\bibitem{weyl1919statischen}
H.~Weyl.
\newblock {{\"U}ber die statischen kugelsymmetrischen L{\"o}sungen von
  Einsteins “kosmologischen” Gravitationsgleichungen}.
\newblock {\em Phys. Z}, 20(31-34):65, 1919.

\bibitem{Islam:1983rxp}
J.~N. Islam.
\newblock {The cosmological constant and classical tests of general
  relativity}.
\newblock {\em Phys. Lett. A}, 97:239--241, 1983.

\bibitem{Rindler:2007zz}
W.~Rindler and M.~Ishak.
\newblock {Contribution of the cosmological constant to the relativistic
  bending of light revisited}.
\newblock {\em Phys. Rev. D}, 76:043006, 2007.

\bibitem{Bhattacharya:2009rv}
A.~Bhattacharya, A.~Panchenko, M.~Scalia, C.~Cattani, and K.~K. Nandi.
\newblock {Light bending in the galactic halo by Rindler-Ishak method}.
\newblock {\em JCAP}, 09:004, 2010.

\bibitem{Bhattacharya:2010xh}
A.~Bhattacharya, G.~M. Garipova, E.~Laserra, A.~Bhadra, and K.~K. Nandi.
\newblock {The Vacuole Model: New Terms in the Second Order Deflection of
  Light}.
\newblock {\em JCAP}, 02:028, 2011.

\bibitem{Hu:2021yzn}
L.~Hu, A.~Heavens, and D.~Bacon.
\newblock {Light bending by the cosmological constant}.
\newblock {\em JCAP}, 02(02):009, 2022.

\bibitem{Bessa:2022sdh}
P.~Bessa and O.~F. Piattella.
\newblock {Gravitational lensing in a universe with matter and a cosmological
  constant}.
\newblock {\em Phys. Rev. D}, 106(12):123513, 2022.

\bibitem{Lake:2001fj}
K.~Lake.
\newblock {Bending of light and the cosmological constant}.
\newblock {\em Phys. Rev. D}, 65:087301, 2002.

\bibitem{Gibbons:2008rj}
G.~W. Gibbons and M.~C. Werner.
\newblock {Applications of the Gauss-Bonnet theorem to gravitational lensing}.
\newblock {\em Class. Quant. Grav.}, 25:235009, 2008.

\bibitem{Werner:2012rc}
M.~C. Werner.
\newblock {Gravitational lensing in the Kerr-Randers optical geometry}.
\newblock {\em Gen. Rel. Grav.}, 44:3047--3057, 2012.

\bibitem{Crisnejo_2018}
G.~Crisnejo and E.~Gallo.
\newblock Weak lensing in a plasma medium and gravitational deflection of
  massive particles using the gauss-bonnet theorem. a unified treatment.
\newblock {\em Phys. Rev. D}, 97:124016, Jun 2018.

\bibitem{Ovgun:2018fnk}
A.~\"Ovg\"un.
\newblock {Light deflection by Damour-Solodukhin wormholes and Gauss-Bonnet
  theorem}.
\newblock {\em Phys. Rev. D}, 98(4):044033, 2018.

\bibitem{Jusifi_2017.2}
K.~Jusufi, I.~Sakalli, and A.~\"Ovg\"un.
\newblock {Effect of Lorentz symmetry breaking on the deflection of light in a
  cosmic string spacetime}.
\newblock {\em Phys. Rev. D}, 96:024040, Jul 2017.

\bibitem{Ishihara:2016vdc}
A.~Ishihara, Y.~Suzuki, T.~Ono, T.~Kitamura, and H.~Asada.
\newblock {Gravitational bending angle of light for finite distance and the
  Gauss-Bonnet theorem}.
\newblock {\em Phys. Rev. D}, 94(8):084015, 2016.

\bibitem{Ishihara:2016sfv}
A.~Ishihara, Y.~Suzuki, T.~Ono, and H.~Asada.
\newblock {Finite-distance corrections to the gravitational bending angle of
  light in the strong deflection limit}.
\newblock {\em Phys. Rev. D}, 95(4):044017, 2017.

\bibitem{Takizawa:2020egm}
K.~Takizawa, T.~Ono, and H.~Asada.
\newblock {Gravitational deflection angle of light: Definition by an observer
  and its application to an asymptotically nonflat spacetime}.
\newblock {\em Phys. Rev. D}, 101(10):104032, 2020.

\bibitem{Takizawa:2020dja}
K.~Takizawa, T.~Ono, and H.~Asada.
\newblock {Gravitational lens without asymptotic flatness: Its application to
  the Weyl gravity}.
\newblock {\em Phys. Rev. D}, 102(6):064060, 2020.

\bibitem{Arakida:2017hrm}
H.~Arakida.
\newblock {Light deflection and Gauss\textendash{}Bonnet theorem: definition of
  total deflection angle and its applications}.
\newblock {\em Gen. Rel. Grav.}, 50(5):48, 2018.

\bibitem{Arakida:2020xil}
H.~Arakida.
\newblock {The optical geometry definition of the total deflection angle of a
  light ray in curved spacetime}.
\newblock {\em JCAP}, 08:028, 2021.

\bibitem{Wald:1984rg}
R.~M. Wald.
\newblock {\em {General Relativity}}.
\newblock Chicago Univ. Pr., Chicago, USA, 1984.

\bibitem{Diff_GeomCarmo}
M.~P. do~Carmo.
\newblock {\em {Differential Geometry of Curves and Surfaces}}.
\newblock Dover Publications, Mineola, New York, 2 edition, 2016.

\bibitem{abbena2006modern}
E.~Abbena, S.~Salamon, and A.~Gray.
\newblock {\em {Modern Differential Geometry of Curves and Surfaces with
  Mathematica. Third Edition}}.
\newblock Textbooks in Mathematics. Taylor \& Francis, 2006.

\bibitem{VPerlick_1990}
V.~Perlick.
\newblock {On Fermat's principle in general relativity. I. The general case}.
\newblock {\em {Class. Quant. Grav.}}, 7(8):1319, 1990.

\bibitem{Ono:2017pie}
T.~Ono, A.~Ishihara, and H.~Asada.
\newblock {Gravitomagnetic bending angle of light with finite-distance
  corrections in stationary axisymmetric spacetimes}.
\newblock {\em Phys. Rev. D}, 96(10):104037, 2017.

\bibitem{oprea2007differentia}
J.~Oprea.
\newblock {\em {Differential Geometry and Its Applications}}.
\newblock Classroom resource materials. Mathematical Association of America,
  2007.

\bibitem{Arakida:2011ty}
H.~Arakida and M.~Kasai.
\newblock {Effect of the cosmological constant on the bending of light and the
  cosmological lens equation}.
\newblock {\em Phys. Rev. D}, 85:023006, 2012.

\bibitem{Roque_On_the_radial_2023}
{Repository}.
\newblock Github.com/Mandy8808/Implementation.git, 2023.

\bibitem{will_2018}
C.~M. Will.
\newblock {\em {Theory and Experiment in Gravitational Physics}}.
\newblock Cambridge University Press, 2 edition, 2018.

\bibitem{Epstein:1980dw}
R.~Epstein and I.~I. Shapiro.
\newblock {Post-post-Newtonian deflection of light by the Sun}.
\newblock {\em Phys. Rev. D}, 22:2947--2949, 1980.

\bibitem{Ishak:2007ea}
M.~Ishak, W.~Rindler, J.~Dossett, J.~Moldenhauer, and C.~Allison.
\newblock {A New Independent Limit on the Cosmological Constant/Dark Energy
  from the Relativistic Bending of Light by Galaxies and Clusters of Galaxies}.
\newblock {\em Mon. Not. Roy. Astron. Soc.}, 388:1279--1283, 2008.

\bibitem{Horndeski:1974wa}
G.~W. Horndeski.
\newblock {Second-order scalar-tensor field equations in a four-dimensional
  space}.
\newblock {\em Int. J. Theor. Phys.}, 10:363--384, 1974.

\bibitem{Deffayet:horndeski}
C.~Deffayet, G.~Esposito-Farese, and A.~Vikman.
\newblock {Covariant Galileon}.
\newblock {\em Phys. Rev. D}, 79:084003, 2009.

\bibitem{Deffayet:horndeski2}
C.~Deffayet, S.~Deser, and G.~Esposito-Farese.
\newblock {Generalized Galileons: All scalar models whose curved background
  extensions maintain second-order field equations and stress-tensors}.
\newblock {\em Phys. Rev. D}, 80:064015, 2009.

\bibitem{Afrin:2021wlj}
M.~Afrin and S.~G. Ghosh.
\newblock {Testing Horndeski Gravity from EHT Observational Results for
  Rotating Black Holes}.
\newblock {\em Astrophys. J.}, 932(1):51, 2022.

\bibitem{Atamurotov:2022slw}
F.~Atamurotov, F.~Sarikulov, A.~Abdujabbarov, and B.~Ahmedov.
\newblock {Gravitational weak lensing by black hole in Horndeski gravity in
  presence of plasma}.
\newblock {\em Eur. Phys. J. Plus}, 137(3):336, 2022.

\bibitem{Babichev:2016rlq}
E.~Babichev, C.~Charmousis, and A.~Leh\'ebel.
\newblock {Black holes and stars in Horndeski theory}.
\newblock {\em Class. Quant. Grav.}, 33(15):154002, 2016.

\bibitem{Planck:2018vyg}
N.~Aghanim et~al.
\newblock {Planck 2018 results. VI. Cosmological parameters}.
\newblock {\em Astron. Astrophys.}, 641:A6, 2020.
\newblock [Erratum: Astron.Astrophys. 652, C4 (2021)].

\bibitem{Javed:2020pyz}
W.~Javed, J.~Abbas, Y.~Kumaran, and A.~\"Ovg\"un.
\newblock {Weak deflection angle by asymptotically flat black holes in
  Horndeski theory using Gauss-Bonnet theorem}.
\newblock {\em Int. J. Geom. Meth. Mod. Phys.}, 18(01):2150003, 2021.

\bibitem{2020SciPy-NMeth}
P.~Virtanen, R.~Gommers, and et~al.
\newblock {SciPy 1.0: Fundamental Algorithms for Scientific Computing in
  Python}.
\newblock {\em Nature Methods}, 17:261--272, 2020.

\bibitem{DORMAND198019}
J.~R. Dormand and P.~J. Prince.
\newblock {A family of embedded Runge-Kutta formulae}.
\newblock {\em Journal of Computational and Applied Mathematics}, 6(1):19--26,
  1980.

\bibitem{Lawrence1986SomePR}
F.~S. Lawrence.
\newblock {Some practical Runge-Kutta formulas}.
\newblock {\em Mathematics of Computation}, 46:135--150, 1986.

\bibitem{Perlick:2021aok}
V.~Perlick and O.~Y. Tsupko.
\newblock {Calculating black hole shadows: Review of analytical studies}.
\newblock {\em Phys. Rept.}, 947:1--39, 2022.

\bibitem{Alfio2007}
A.~Quarteroni, R.~Sacco, and F.~Saleri.
\newblock {\em {Numerical Mathematics}}.
\newblock Springer New York, NY, 2007.

\bibitem{Turyshev:2003wt}
S.~G. Turyshev, M.~Shao, and K.~Nordtvedt, Jr.
\newblock {The Laser astrometric test of relativity mission}.
\newblock {\em Class. Quant. Grav.}, 21:2773--2799, 2004.

\bibitem{Sereno:2007rm}
M.~Sereno.
\newblock {On the influence of the cosmological constant on gravitational
  lensing in small systems}.
\newblock {\em Phys. Rev. D}, 77:043004, 2008.

\bibitem{Sereno:2008kk}
M.~Sereno.
\newblock {The role of Lambda in the cosmological lens equation}.
\newblock {\em Phys. Rev. Lett.}, 102:021301, 2009.

\bibitem{Synge:1966okc}
J.~L. Synge.
\newblock {The Escape of Photons from Gravitationally Intense Stars}.
\newblock {\em Mon. Not. Roy. Astron. Soc.}, 131(3):463--466, 1966.

\bibitem{Qiao:2022jlu}
C.~K. Qiao and M.~Li.
\newblock {Geometric approach to circular photon orbits and black hole
  shadows}.
\newblock {\em Phys. Rev. D}, 106(2):L021501, 2022.

\bibitem{1983BAICz..34..129S}
Z.~{Stuchlik}.
\newblock {The Motion of Test Particles in Black-Hole Backgrounds with Non-Zero
  Cosmological Constant}.
\newblock {\em Bulletin of the Astronomical Institutes of Czechoslovakia},
  34:129, June 1983.

\bibitem{PhysRevD.60.044006}
Z.~Stuchl\'{\i}k and S.~Hled\'{\i}k.
\newblock {Some properties of the Schwarzschild--de Sitter and
  Schwarzschild--anti-de Sitter spacetimes}.
\newblock {\em Phys. Rev. D}, 60:044006, Jul 1999.

\bibitem{Perlick:2018iye}
V.~Perlick, O.~Y. Tsupko, and G.~S. Bisnovatyi-Kogan.
\newblock {Black hole shadow in an expanding universe with a cosmological
  constant}.
\newblock {\em Phys. Rev. D}, 97(10):104062, 2018.

\bibitem{Roy:2020dyy}
R.~Roy and S.~Chakrabarti.
\newblock {Study on black hole shadows in asymptotically de Sitter spacetimes}.
\newblock {\em Phys. Rev. D}, 102(2):024059, 2020.

\bibitem{Tsupko:2019mfo}
O.~Y. Tsupko and G.~S. Bisnovatyi-Kogan.
\newblock {First analytical calculation of black hole shadow in McVittie
  metric}.
\newblock {\em Int. J. Mod. Phys. D}, 29(09):2050062, 2020.

\bibitem{Chandrasekhar:1985kt}
S.~Chandrasekhar.
\newblock {\em The Mathematical Theory of Black Holes}.
\newblock International series of monographs on physics. Clarendon Press, 1998.

\bibitem{Grenzebach_2015}
A.~Grenzebach, V.~Perlick, and C.~L\"ammerzahl.
\newblock {Photon regions and shadows of accelerated black holes}.
\newblock {\em International Journal of Modern Physics D}, 24(09):1542024,
  2015.

\bibitem{Bisnovatyi-Kogan:2018vxl}
G.~S. Bisnovatyi-Kogan and O.~Y. Tsupko.
\newblock {Shadow of a black hole at cosmological distances}.
\newblock {\em Phys. Rev. D}, 98(8):084020, 2018.

\bibitem{Bellini:2015xja}
E.~Bellini, A.~J. Cuesta, R.~Jimenez, and L.~Verde.
\newblock {Constraints on deviations from \ensuremath{\Lambda}CDM within
  Horndeski gravity}.
\newblock {\em JCAP}, 02:053, 2016.
\newblock [Erratum: JCAP 06, E01 (2016)].

\bibitem{Noller:2018wyv}
J.~Noller and A.~Nicola.
\newblock {Cosmological parameter constraints for Horndeski scalar-tensor
  gravity}.
\newblock {\em Phys. Rev. D}, 99(10):103502, 2019.

\bibitem{Ni:2012eh}
W.~T. Ni.
\newblock {ASTROD-GW: Overview and Progress}.
\newblock {\em Int. J. Mod. Phys. D}, 22:1341004, 2013.

\bibitem{Bayle:2022hvs}
J.~B. Bayle, B.~Bonga, C.~Caprini, D.~Doneva, M.~Muratore, A.~Petiteau,
  E.~Rossi, and L.~Shao.
\newblock {Overview and progress on the Laser Interferometer Space Antenna
  mission}.
\newblock {\em Nature Astron.}, 6(12):1334--1338, 2022.

\bibitem{Amaro_Seoane_2023}
P.~Amaro-Seoane, J.~Andrews, M.~Arca-Sedda, A.~Askar, and et~al.
\newblock Astrophysics with the laser interferometer space antenna.
\newblock {\em Living Reviews in Relativity}, 26(1), mar 2023.

\bibitem{Ni:2016wcv}
W.~T. Ni.
\newblock {Gravitational wave detection in space}.
\newblock {\em Int. J. Mod. Phys. D}, 25(14):1630001, 2016.

\bibitem{PhysRevD.97.124016}
G.~Crisnejo and E.~Gallo.
\newblock Weak lensing in a plasma medium and gravitational deflection of
  massive particles using the gauss-bonnet theorem. a unified treatment.
\newblock {\em Phys. Rev. D}, 97:124016, Jun 2018.

\end{thebibliography}
\end{document}